\documentclass[usenatbib]{mnras}
\pdfoutput=1
\usepackage[latin1]{inputenc}
\usepackage{graphicx}
\usepackage{amsmath}
\usepackage{amsfonts}
\usepackage{amssymb}
\usepackage{siunitx}
\usepackage{epstopdf}

\title{A local model of warped magnetised accretion discs}

\author[Paris and Ogilvie]{
J. B. Paris$^{1}$
and G. I. Ogilvie$^{1}$
\\
$^{1}$Department of Applied Mathematics and Theoretical Physics, Centre for Mathematical Sciences, University of Cambridge,\\
 Wilberforce Road, Cambridge, CB3 0WA}

\begin{document}
\maketitle

\begin{abstract}
We derive expressions for the local ideal magnetohydrodynamic (MHD) equations for a warped astrophysical disc using a warped shearing box formalism. A perturbation expansion of these equations to first order in the warping amplitude leads to a linear theory for the internal local structure of magnetised warped discs in the absence of MRI turbulence. In the special case of an external magnetic field oriented normal to the disc surface these equations are solved semi-analytically via a spectral method. The relatively rapid warp propagation of low-viscosity Keplerian hydrodynamic warped discs is diminished by the presence of a magnetic field. The magnetic tension adds a stiffness to the epicyclic oscillations, detuning the natural frequency from the orbital frequency and thereby removing the resonant forcing of epicyclic modes characteristic of hydrodynamic warped discs. In contrast to a single hydrodynamic resonance we find a series of Alfv\'{e}nic-epicyclic modes which may be resonantly forced by the warped geometry at critical values of the orbital shear rate $q$ and magnetic field strength. At these critical points large internal torques are generated and anomalously rapid warp propagation occurs. As our treatment omits MRI turbulence, these results are of greatest applicability to strongly magnetised discs.\\
\end{abstract}

\section{Introduction}
	Warped astrophysical discs are discs in which the orientation of the orbital plane is dependent on the radius. These warps may be produced by a wide variety of sources, such as Lense-Thirring torques \citep{1975ApJ...195L..65B} from a central misaligned spinning black hole or the torques due to a misaligned companion object in a binary star system \citep{2000A&A...360.1031T}. Theoretical considerations and observations revealing the existence of warped discs in several systems, such as the X-ray binary Her X-1 and the galactic nucleus of NGC 4258 (M106) \citep{1995Natur.373..127M}, have historically motivated the study of warped accretion disc dynamics.\\
	 
	Early theoretical studies \citep{1975ApJ...195L..65B} \citep{1981ApJ...247..677H} found evolutionary equations for the shape of a warped disc and suggested that the warp would diffuse on a viscous time-scale inversely proportional to $\alpha$, where $\alpha$ is the Shakura-Sunyaev viscosity parameter. These models were challenged by \cite{1983MNRAS.202.1181P} who determined that the prior models failed to appropriately recognize the importance of internal flows within the disc and consequently did not conserve angular momentum. \\
	
	 \cite{1995ApJ...438..841P} and earlier works showed that warps in inviscid Keplerian discs propagate as bending waves at a fraction of the sound speed. The fully non-linear theory for the diffusive regime in Keplerian discs and for the slower bending waves in non-Keplerian discs was derived by \cite{1999MNRAS.304..557O}. In that work it was shown that warped hydrodynamic discs exhibit resonant behaviour when the disc is Keplerian and the warp is small.  \\

More recently, \cite{2013MNRAS.433.2403O} have shown that the nonlinear hydrodynamic theory of Ogilvie (1999) can be derived by separating the global and local aspects of a warped thin disc.  The global evolution of the mass distribution and the shape of the disc can be deduced from the (vectorial) conservation of angular momentum, provided that the internal torque is known.  The internal torque can be calculated from time-averaged quantities in a local model, which is constructed around a reference point that follows a circular orbit through a warped disc and experiences a geometry that oscillates at the orbital frequency, thereby generalizing the well known shearing sheet  or box.  The simplest solutions in the warped shearing box are laminar internal flows that oscillate at the orbital frequency; these are driven by a radial pressure gradient that arises from the warped geometry and the vertical stratification of the disc.  Owing to the coincidence of the orbital and epicyclic frequency in a Keplerian disc, this forcing may result in fast internal flows for even comparatively small warps. Such flows generate correspondingly large internal torques, leading to the rapid propagation of the warp.\\

	There are many situations in which these hydrodynamic models are not valid. Astrophysical discs are often threaded by large or small-scale magnetic fields thought to be important in the formation of jet outflows \citep{1982MNRAS.199..883B} and the transport of angular momentum via the magnetorotational instability \citep{1991ApJ...376..214B}. These magnetised discs may be created, for example, by dragging in significant amounts of magnetic flux during disc formation as is thought to be the case for protostellar discs \citep{2017MNRAS.467.3324L}. It is therefore likely that there exist many disc systems in which both large-scale magnetic fields and the warped geometry will have important effects on disc dynamics. Despite this, the study of warped discs and the study of magnetised discs have generally been done in relative isolation. With a few notable exceptions, comparatively little attention has been given to the complex interplay between these two aspects of accretion disc physics.\\
	
	From a computational perspective it has been difficult to resolve warp evolution in large-scale MHD simulations. This difficulty lies in the vast separation between the timescale of MHD turbulence, which requires a time-step very short relative to the orbital period, and the warp evolution timescale which is usually far longer than the orbital period. However, a limited body of work on warped magnetised discs does exist. \cite{2007ApJ...668..417F} performed one of the first 3D MHD simulations of a relatively thick tilted disc around a rotating black hole. The radial extent of the simulation was unfortunately not large enough to capture significant disc warping. The simulations of tilted discs around stellar objects of \cite{2014EPJWC..6405003L} and \cite{2015ApJ...814..113S} on the other hand were concerned primarily with Rossby wave trapping and jet launching respectively, providing relatively little insight on the warp evolution process. \\

	The global 3D MHD simulations of a warped disc recently performed by \cite{2013ApJ...777...21S} were among the first to resolve shape evolution in a magnetised accretion disc. In that work it was argued that neither the bending wave nor the diffusion models of warp propagation were adequate to properly describe the observed warp propagation. Qualitative differences between the hydrodynamic and MHD warped disc simulations (most notably a distinctively sharper warp profile in the hydrodynamic case) were also observed. Although the departure from standard hydrodynamic warped disc theory was partially explained through non-linear damping effects, the magnetohydrodynamic aspect was clearly recognised and arguments involving the MRI-driven background turbulence and anisotropic viscosities were presented to rationalise the observed warp propagation. These results raise questions about the applicability of certain aspects of hydrodynamic warped disc theory (especially the neat separation of discs into `diffusive' or `bending wave' regimes) to magnetised discs. A contrary view has been presented by \cite{2016MNRAS.455L..62N}, who found that the results of a similar MHD simulation by \cite{2015ApJ...806..141K} -- which does however begin with a weaker mean magnetic field -- could be quite accurately reproduced using a hydrodynamic simulation with an alpha viscosity. Continued interest in Bardeen-Petterson alignment has stimulated the production of a number of recent warped magnetised disc simulations such as \cite{2014ApJ...796..104Z}. The advent of 3D MHD simulations of this kind necessitates the development a corresponding theory of warp propagation in magnetised discs in order to explain the salient features of these simulations. \\
	
	Bending instabilities in a magnetised warped disc subject to an inclined dipole field were studied in a vertically integrated manner by \cite{1997MNRAS.292..631A}. \cite{1999ApJ...524.1030L} investigated warping in a magnetised disc due to an inclined dipole, but did not include the effects of the warped disc geometry itself on the disc response. \cite{2000A&A...360.1031T} included this effect and were able to determine the global warped disc structure for a variety of magnetic field configurations. However, none of these models considered the potentially dramatic effects of driven internal flows such as those described by \cite{2013MNRAS.433.2403O} in hydrodynamic warped discs. The work presented in this paper attempts to analyse whether these dramatic resonantly driven flows and their associated large torques exist in magnetised warped discs, and if so, what consequences they would have for the evolution of the warp. \\
	
		The problem of a magnetised warped disc presents a number of different issues and regimes, only some of which can be addressed in this paper.  If the disc is threaded by a large-scale magnetic field, the warping of the disc will affect the structure of this external field and will generally produce torques on the disc that will cause the warp to evolve.  This aspect of the problem is global in character and can be studied only to a limited extent using the local model considered in this paper.  A magnetic field will also affect the interior dynamics of the disc.  If it is sufficiently weak, it can cause the magnetorotational instability, which can produce MHD turbulence.  The occurrence of the MRI in a warped disc, its effect on the internal flows and torques, and its interaction with the hydrodynamic instability studied by \cite{2013MNRAS.433.2420O} could be usefully studied in the local model of the warped shearing box.  The launching of outflows from a magnetized disc can be studied to some extent using local models \citep{2012MNRAS.423.1318O} and the interaction of that problem with a warp also deserves investigation.  In this paper we take a useful first step by calculating the laminar internal flows in a warped disc threaded by a magnetic field in ideal MHD.  The laminar solutions are of most direct relevance when the field is sufficiently strong to suppress the MRI.\\
	
	In section 2 we employ the warped shearing box model developed by \cite{2013MNRAS.433.2403O} to express the fully non-linear local ideal MHD equations in a reference frame comoving with a fluid element in the warped disc midplane. In section 3 we consider one-dimensional laminar solutions to these equations, and by expanding linearly in warp amplitude about an equilibrium solution, a system of first-order differential equations is found for the internal structure of the disc.  \\	
	
	In section 4 we consider the special case of a warped disc threaded by a magnetic field perpendicular to the disc midplane. A solution to this particular case is found via a spectral method. The unique features of magnetised warped disc dynamics are discussed. In section 5 we outline two numerical procedures that are used to check the results of the spectral method. In section 6 the results of the paper are discussed and summarized.\\

\section{The local model of a warped disc}

\subsection{The warped shearing box}

We restrict ourselves to circular warped discs (as opposed to elliptical discs) due to their relative simplicity. \cite{1999MNRAS.304..557O} showed that a circular warped disc may be mathematically described by a radially dependent vector $\mathbf{l}(r,t)$ defined to be perpendicular to the orbital plane of the fluid at radius $r$. The radial dependence of the orbital plane causes a warping in the disc geometry, as can be seen in Figure 1.  For this class of accretion disc, only the dimensionless warp amplitude $\psi$ defined by $\psi = r \vert \frac{\partial \mathbf{l}}{\partial r} \vert$ is necessary to specify the local geometry of the warped disc. The physical significance of the warping parameter $\psi$ can be clearly seen in Figure 2\footnote{Figure 1 and Figure 2 are derived from figures originally published in \cite{2013MNRAS.433.2403O}}. \\

Local models of unwarped accretion discs are typically constructed around a reference point in the midplane of the disc co-orbiting with the surrounding fluid. A Cartesian coordinate system can be imposed with the origin at this reference point and the axes rotating at the orbital frequency such that the $x$-direction is radial, the $y$-direction is azimuthal and the $z$-direction is perpendicular to the disc midplane. There is a shearing flow within the box due the differential rotation of the disc; consequently such models are often called `shearing box' models.\\

An analogous construction is the `warped shearing box', introduced in \cite{2013MNRAS.433.2403O}. In warped shearing box models, a Cartesian coordinate system is constructed around a reference point in the mid-plane of a warped accretion disc as shown in Figure 1. The warped shearing box is represented by the non-orthogonal primed coordinates $(t',x',y',z')$ chosen to compensate for the local geometry of the warped disc. These are related to the Cartesian coordinate system $(t,x,y,z)$ comoving with the fluid via expressions \eqref{coord1}--\eqref{coord4} below, where $q\equiv -\frac{d \ln \Omega}{d\ln r}$ is the orbital shear rate of the disc in question and $\Omega$ is the orbital frequency at the reference radius.\\

We define $z'$ such that the surface $z'=0$ corresponds to the midplane of the warped disc and the $z'$-axis is normal to the disc midplane rather than normal to the orbital plane. The definition of $y'$ simply takes into account the shear due to the differential rotation, where $q$ is assumed to be 3/2 for a Keplerian disc. The relative velocities $\mathbf{v}$ are constructed to remove the expected azimuthal velocity changes due to shear or the vertical velocity changes due to the warped geometry of the disc.

\begin{figure}
  \centering
    \includegraphics[width=0.5\textwidth]{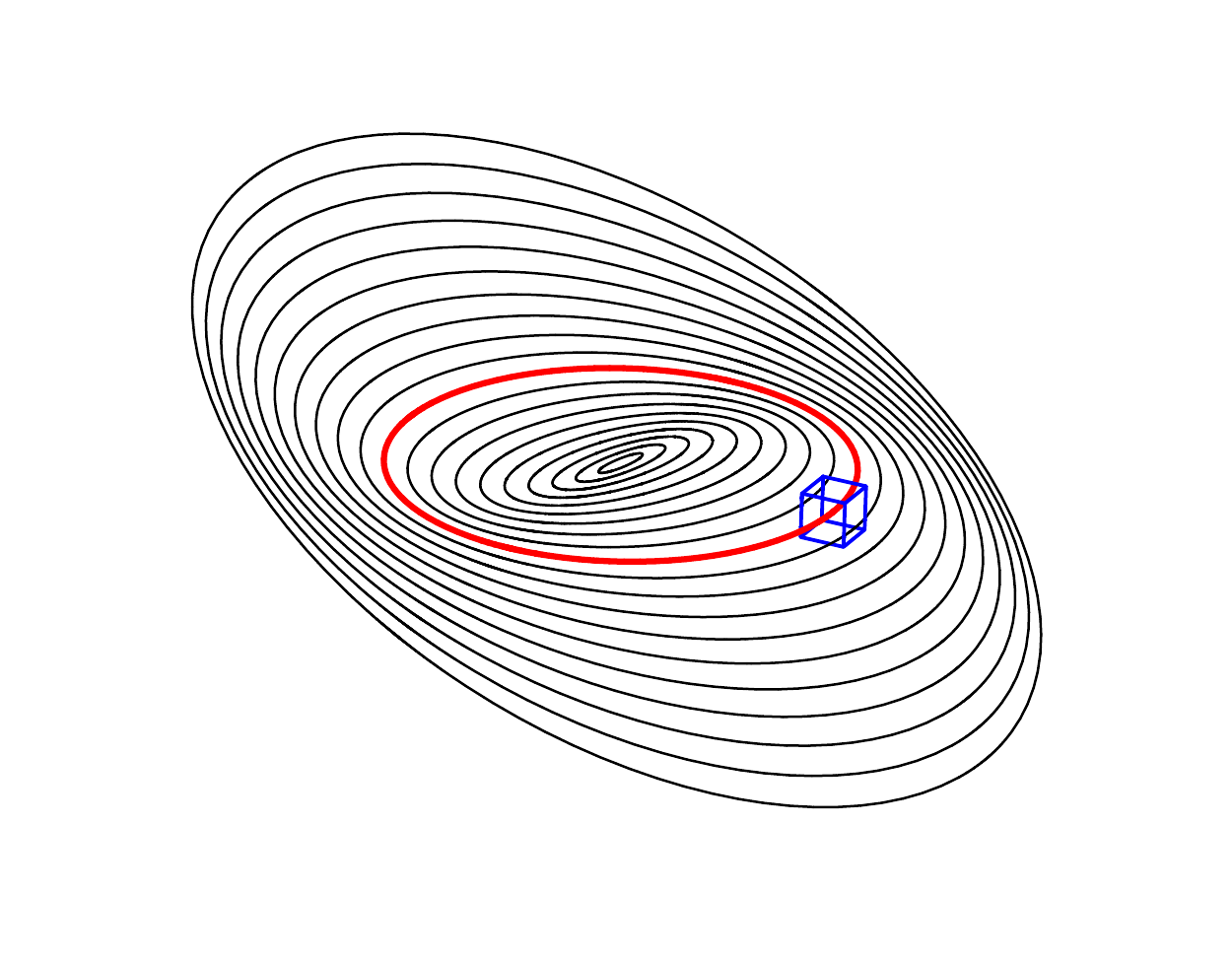}
     \caption{A warped disc, viewed as a collection of tilted rings. As discussed in Section 2.1, a local frame (the blue box) is centred on a point that follows the red reference orbit.  }
\end{figure}

\begin{figure}
  \centering
    \includegraphics[width=0.4\textwidth]{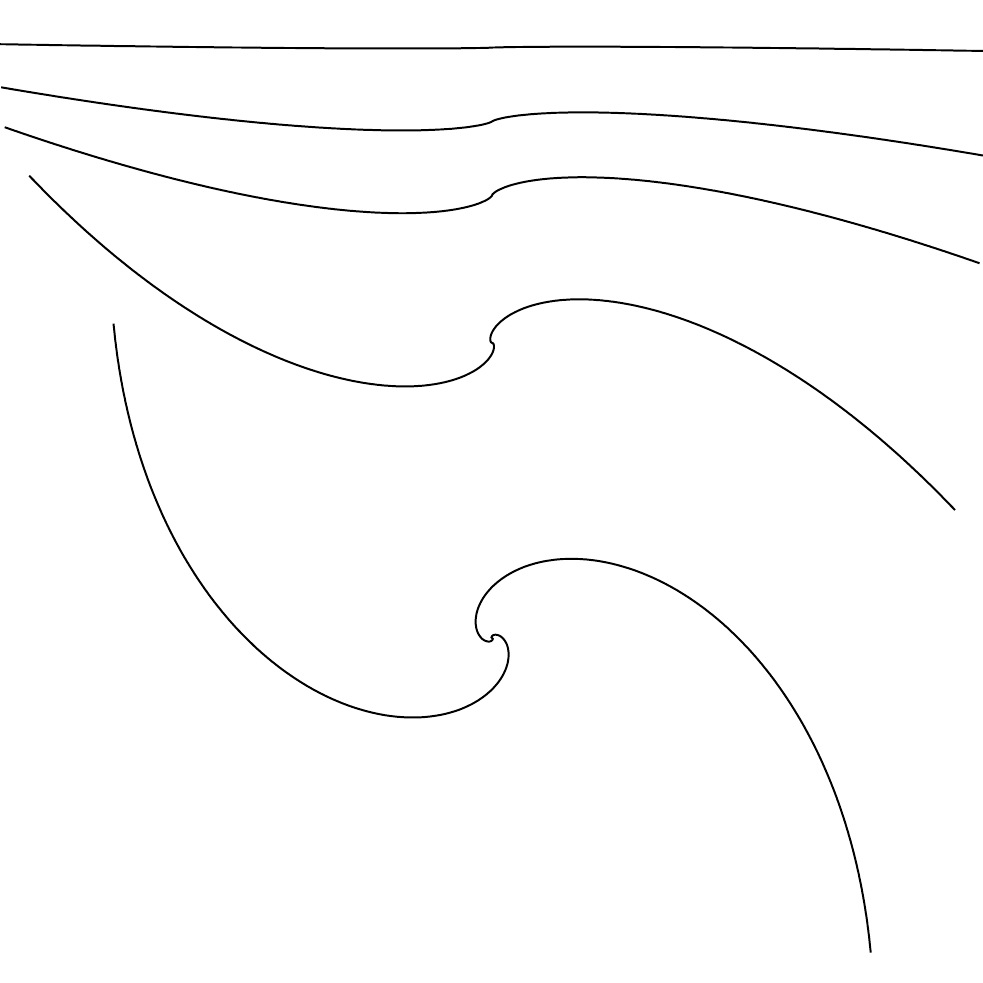}
     \caption{A side-on view of untwisted warps with constant warp amplitudes $ \psi = 0.01$(top), $0.1$, $0.2$, $0.5$, and $1$(bottom).}
\end{figure}	

\begin{equation}
t'=t,
\label{coord1}
\end{equation}

\begin{equation}
x'=x,
\label{coord2}
\end{equation}

\begin{equation}
y'=y + q \Omega t x,
\label{coord3}
\end{equation}

\begin{equation}
z'=z + \psi x \cos(\Omega t).
\label{coord4}
\end{equation}
In such coordinates, it is important to note that

\begin{equation}
\partial _t = \partial _{t'} + q \Omega x \partial _{y'} - \psi \Omega x \sin(\Omega t)  \partial _{z'}, 
\end{equation}

\begin{equation}
\partial _x = \partial _{x'} + q \Omega t \partial _{y '} + \psi \cos(\Omega t) \partial _{z'},
\end{equation}

\begin{equation}
\partial _y = \partial _{y'},
\end{equation}

\begin{equation}
\partial _z = \partial _{z'},
\end{equation}
and relative velocities are defined such that

\begin{equation}
v_x = u_x,
\end{equation}

\begin{equation}
v_y = u_y + q \Omega x,
\end{equation}

\begin{equation}
v_z = u_z - \psi \Omega x \sin (\Omega t).
\end{equation}

\subsection{The non-linear MHD equations in a warped disc}

Having defined the coordinate system of the warped shearing box, we may now express the full non-linear local MHD equations in these coordinates.\\

\subsubsection{The physical equations}

The momentum equation expressed in a coordinate system corotating with the reference orbit is

\begin{equation}
\rho \left( \frac{D\mathbf{u}}{Dt} + 2 \Omega \mathbf{e}_z \wedge \mathbf{u} \right) = -\rho \nabla \Phi - \nabla p + \mathbf{J} \wedge \mathbf{B},
\end{equation}	
where the effective potential due to a central gravitational force and the rotating frame is given by

\begin{equation}
\Phi = \frac{1}{2} \Omega ^2 (z^2-2q x^2)
\end{equation}
and the Lorentz force in ideal MHD is expressed as

\begin{equation}
\mathbf{J} \wedge \mathbf{B} = \frac{1}{\mu _0}(\mathbf{B} \cdot \nabla) \mathbf{B} - \frac{1}{2 \mu_0} \nabla \mathbf{B}^2.
\end{equation}
The induction equation in ideal MHD is given by

\begin{equation}
\frac{\partial \mathbf{B}}{\partial t} = \nabla \wedge (\mathbf{u} \wedge \mathbf{B}) = \mathbf{B} \cdot \nabla\mathbf{u} - \mathbf{u} \cdot \nabla \mathbf{B} - \mathbf{B} (\nabla \cdot \mathbf{u}),
\end{equation}
while the continuity equation gives
	
	\begin{equation}
\frac{\partial \rho}{\partial t} + \nabla \cdot (\rho \mathbf{u}) = 0.
\end{equation} 
For simplicity the disc is assumed to be isothermal with uniform sound speed $c_s$, implying
	
\begin{equation}
p = c_s ^2 \rho.
\end{equation} 

Rather than expressing the full MHD equations in terms of the vector components perpendicular to the orbital plane, such as $v_{z}$ and $B_{z}$, it will be more convenient to consider the vector components perpendicular to the disc midplane as defined by $v_{z'} = v_{z} + \psi \cos(\Omega t) v_{x}$, $B_{z'} = B_{z} + \psi \cos(\Omega t) B_{x}$ (in fact, these are contravarient vector components in the warped shearing coordinate system). The full non-linear MHD equations are then given by

\begin{equation}
\begin{split}
\frac{D v_x}{Dt'} - 2 \Omega v_y = -\frac{1}{\rho}[\partial_{x'} + q \Omega t' \partial_{y'} + \psi \cos(\Omega t')\partial_{z'} ]\Pi
\\+ \frac{1}{\mu_0 \rho} (\mathbf{B' \cdot \nabla}) B_{x},  
\end{split} \label{nonlin1}
\end{equation}

\begin{equation}
\begin{split}
\frac{D v_y}{Dt'} + (2-q) \Omega v_{x}  = -\frac{1}{\rho} \partial_{y'}\Pi 
+ \frac{1}{ \mu_0 \rho} (\mathbf{B' \cdot \nabla}) B_{y},
\end{split}
\end{equation}

 \begin{equation}
\begin{split}
 \frac{D v_{z'}}{Dt'} = -2 \psi \Omega \sin(\Omega t') v_x   -  \Omega^2 z'  + 2 \Omega \psi \cos(\Omega t')  v_y \\
 - \frac{\gamma^2 }{\rho} \partial_{z'} \Pi 
-\frac{1}{\rho} \psi \cos(\Omega t') (\partial_{x'}  + q \Omega t' \partial_{y'} )\Pi +
 \frac{1}{\mu_0 \rho}(\mathbf{B'
  \cdot \nabla}) B_{z'},
  \end{split}
 \end{equation}

\begin{equation}
\frac{\partial B_x}{\partial t'} =   \mathbf{B'} \cdot \nabla v_{x} -  (\nabla \cdot \mathbf{v'}) B_{x} - \mathbf{v'} \cdot \nabla B_{x},
\end{equation} 

 \begin{equation}
 \begin{split}
\frac{\partial B_y}{\partial t'} = - q \Omega B_x + \mathbf{B'} \cdot \nabla v_{y} -  (\nabla \cdot \mathbf{v'}) B_{y} - \mathbf{v'} \cdot \nabla B_{y},
 \end{split}
\end{equation} 

\begin{equation}
\begin{split}
\frac{\partial B_{z'}}{\partial t'} =\mathbf{B'} \cdot \nabla v_{z'} -  (\nabla \cdot \mathbf{v'}) B_{z'} - \mathbf{v'} \cdot \nabla B_{z'},
\end{split}
\end{equation}

\begin{equation}
 \nabla \cdot \mathbf{B'}  =  0,
\end{equation}
 
 \begin{equation}
 \partial_{t'} \rho + \rho (\nabla \cdot\mathbf{v'}) + \mathbf{v'} \cdot \nabla \rho=0,
 \end{equation}

where

 \begin{equation}
\frac{D}{Dt'} = \partial_{t'} + \mathbf{v'} \cdot \nabla,
 \end{equation}

\begin{equation}
\mathbf{v'} \cdot \nabla = v_{x} \partial_{x'} + (v_{y} + q \Omega t' v_{x}) \partial_{y'} + v_{z'} \partial_{z'},
\end{equation}

\begin{equation}
 \nabla \cdot \mathbf{v'}  =  \partial_{x'} v_{x}  +  \partial_{y'} (v_{y} + q \Omega t' v_{x}) +  \partial_{z'} v_{z'},
\end{equation}

\begin{equation}
\mathbf{B'} \cdot \nabla = B_{x} \partial_{x'} + (B_{y} + q \Omega t' B_{x}) \partial_{y'} + B_{z'} \partial_{z'},
\end{equation}

\begin{equation}
 \nabla \cdot \mathbf{B'}  =  \partial_{x'} B_{x}  +  \partial_{y'} (B_{y} + q \Omega t' B_{x}) +  \partial_{z'} B_{z'}, \label{nonlin2}
\end{equation}

total pressure $\Pi \equiv c_s ^2 \rho + \frac{\mathbf{B}^2}{2 \mu_0}$ and $\gamma^2 \equiv 1+\psi^2 \cos^2(\Omega t')$.\\

This system of equations is horizontal homogeneous as it does not refer explicitly to $x'$ or $y'$, and so is compatible with periodic horizontal boundary conditions. In order to solve this system of equations, boundary conditions far from the disc midplane must be imposed. More specifically limiting magnetic field values, along with a condition on material inflow/outflow, must be prescribed.\\
	
\subsubsection{The validity of the local model}	
	
It is imperative that we keep track of the assumptions made to derive these equations. Firstly, we have assumed an isothermal gas for simplicity. Perhaps more subtly there is also an implicit separation of scales in the formulation of these equations. The warp, expressed by the warp amplitude $\psi$, is assumed to be independent of time. This reflects the assumption that the warp propagates on a timescale far longer than the orbital timescale ($\vert \frac{d \psi}{dt} \vert \ll \Omega \psi$).\\

At first glance it may seem curious that the warp amplitude is a fixed parameter given that this model is intended partly for the study of warp propagation. However, there is no contradiction here. The warp amplitude determines the internal dynamics of the disc on the fast orbital timescale. An internal torque is generated by the internal flows driven by the warp(as discussed in section 4.2), which can be calculated. In principle the warp could then be allowed to vary on a slower timescale in a manner consistent with the torques acting upon that annulus of the disc. In practice, however, we can deduce the evolution of the warp from the calculated internal torque and the conservation of angular momentum. In this way, warp propagation may be investigated in the local model without an explicitly time-dependent warp amplitude.\\

 \cite{2013MNRAS.433.2403O} displayed that the hydrodynamic warped shearing box is indeed capable of replicating the results of the global asymptotic theory of warped discs \citep{1999MNRAS.304..557O}, validating the local approach. An investigation further exploring the connection between the global and local theory of warped discs is forthcoming. \\

Other challenges may be levelled at this approach to warped disc dynamics. If one were to investigate solutions to these equations that are either one-dimensional or have periodic radial boundaries, one would require the structure of the warp to vary on a length scale much greater than the radial excursion of fluid elements over the period of an orbit. If this were not the case, then the radial variation of the warp amplitude would affect the internal structure of the disc and it would be impossible to characterise a region of the disc by a single value for the warp amplitude. Hence the warped shearing box would be unable to appropriately model a disc with extreme radial variations in warp amplitude. If the warp varies on the length scale of the disc radius this condition simplifies to $\vert \mathbf{v} \vert \ll \Omega R$, or equivalently the relative velocity must be much less than the orbital velocity. \\

Additionally, the presence of a global magnetic field implies that in some sense the transport of angular momentum may not be truly local as is the case for hydrodynamic discs, but rather angular momentum may be communicated non-locally via magnetic stresses to other parts of the disc. This issue enters the local model via several unspecified boundary terms (see section 3.2.2). This topic is discussed in greater detail in the follow-up paper. \\

\section{One dimensional solutions}

\subsection{Non-linear 1D MHD equations}

Having derived the fully non-linear local MHD equations we proceed by looking for the simplest solutions for the disc structure. The warped shearing box was designed such that the MHD equations do not explicitly contain $x'$ or $y'$, and are hence horizontally homogeneous.  We therefore start by seeking solutions that are independent of $x'$ and $y'$. All physical quantities are taken to be functions of $z'$ and $t'$ alone, and the MHD equations reduce to a set of 1D equations. Recall that the oscillatory time-dependence of quantities in the local model corresponds to their azimuthal dependence in a global description (see Fig. 1). We note that such a laminar flow excludes the possibility of turbulence, most notably MRI turbulence.\\

\subsubsection{Scaling}

$B_{z'}$, as shown in the following section, is a constant of the motion and therefore can be used as a scaling parameter. Using this convention the above set of equations \eqref{nonlin1}--\eqref{nonlin2} can be made dimensionless via the following scaling relations:

\begin{equation}
\widehat{t} = \Omega t,
\end{equation}

\begin{equation}
\widehat{z}' = \frac{z' \Omega}{c_s},
\end{equation}

\begin{equation}
\widehat{v}_{i} = v_{i}/c_s,
\end{equation}

\begin{equation}
\widehat{\rho} = \frac{\rho c_s}{\Sigma \Omega},
\end{equation}

\begin{equation}
\widehat{B}_{z'} = \frac{B_{z'}}{\sqrt{\mu_0 \Sigma c_s \Omega}},
\end{equation}

\begin{equation}
\widehat{B}_x = \frac{B_x}{B_{z'}},
\end{equation}
where $c_s$ is the isothermal sound speed, $\Sigma$ is the vertically integrated surface density of the disc, and all other terms maintain their previous definitions. \\

 From this point onwards all physical quantities will be scaled as shown above and the circumflexes will be removed. The one dimensional non-linear MHD equations are hence given by:

\begin{equation}
\begin{split}
\frac{Dv_x}{Dt'} - 2  v_y = - \frac{\psi \cos(t')}{\rho} \partial_{z'} \rho
+ \frac{B_{z'}^2}{\rho} \lbrace  \gamma^2[  \partial_{z'} B_x  \\-
\psi \cos (t') B_x \partial_{z'} B_x] - \psi \cos (t') B_y \partial_{z'} B_y \rbrace,\label{1dnonlin1}
\end{split}
\end{equation}

\begin{equation}
\frac{Dv_y}{Dt'}  = -(2-q) v_x + \frac{B_{z'}^2}{\rho} \partial_{z'}  B_y ,
\end{equation}

 \begin{equation}
\begin{split}
\frac{Dv_{z '}}{Dt'}  = -2 \psi \sin(t') v_x   -  z'  + 2 \psi \cos(t')  v_y - \frac{\gamma^2}{\rho} \partial_{z'} \rho \\
  +\frac{B_{z'}^2 \gamma^2}{\rho } [ - \gamma^2 B_x \partial_{z'} B_x +  \psi \cos(t') \partial_{z'} B_x -  B_y \partial_{z'} B_y ],
  \end{split}
 \end{equation}

 \begin{equation}
\frac{\partial \rho}{\partial t'} +  \partial_{z'} (\rho v_{z'}) = 0,
\end{equation} 

\begin{equation}
\frac{\partial B_x}{\partial t'} =   \partial_{z'} v_x
- v_{z'} \partial_{z'} B_x
 - B_x  \partial_{z'} v_{z'},
\end{equation}

 \begin{equation}
\frac{\partial B_y}{\partial t'} = \partial_{z'} v_y - q B_x
- v_{z'} \partial_{z'} B_y 
 - B_y  \partial_{z'} v_{z'},
\end{equation} 

 \begin{equation}
\frac{\partial B_{z'}}{\partial t'} =0, \label{1dinteresting}
\end{equation} 

 \begin{equation}
\frac{\partial B_{z'}}{\partial z'} =0,\label{1dnonlin2}
\end{equation} 
where $\frac{D}{Dt' } = \partial_{t'} + v_{z'} \partial_{z'} $ and $\gamma^2 \equiv 1+\psi^2 \cos^2(t')$.\\

\subsubsection{Properties of the non-linear 1D MHD equations}

Some general comments can be made at this stage. The final two equations \eqref{1dinteresting}--\eqref{1dnonlin2} reveal an important property of warped magnetized accretion discs. The magnetic field component perpendicular to the disc midplane, $B_{z'}$, is independent of both time and height above the midplane. This is a consequence of magnetic flux conservation under the assumption of horizontal uniformity. The magnetic field perpendicular to the disc midplane is therefore advected with the fluid and a constant of the motion, while the other magnetic field components are not. Thus the use of $B_{z'}$ as a scaling parameter is justified.\\

In a standard unwarped accretion disc there is an azimuthal symmetry present in the equations of motion. With suitable boundary conditions and in the absence of a symmetry-breaking process, there is therefore an azimuthal symmetry in all physical quantities. Although this axisymmetry is broken in a warped disk, equations \eqref{1dnonlin1}--\eqref{1dnonlin2} reveal that a point symmetry still exists. This point symmetry could be described as a reflection in the local disc midplane followed by a rotation by $\pi$ radians. For example, for a scalar quantity $f(z', t')$ this transformation can be represented by the operator $\widehat{T}$ such that

\begin{equation}
\widehat{T} f(z',t') = f(-z', t' + \pi).
\end{equation}

For vector quantities, the vector components parallel to the disc plane are invariant under transformation $\widehat{T}$ and the vector components perpendicular to the disc surface are inverted. For pseudovectors, the opposite is true; pseudovector components parallel to the disc surface are inverted, while pseudovector components perpendicular to the disc surface are invariant. \\

This symmetry can be proven by making the appropriate simultaneous transformation in the warped disc MHD equations: $\lbrace z' \rightarrow -z', t' \rightarrow t' + \pi, \rho \rightarrow \rho, v_{x} \rightarrow v_{x}, v_{y} \rightarrow v_{y}, v_{z'} \rightarrow -v_{z'}, B_{x} \rightarrow -B_{x}, B_{y} \rightarrow -B_{y}, B_{z'} \rightarrow B_{z'} \rbrace$. This transformation leaves equations \eqref{1dnonlin1}--\eqref{1dnonlin2} invariant, and is therefore a symmetry of the system. The existence of this symmetry is not only of physical interest but is of practical use in the numerical work described in sections 4 and 5.\\

\subsection{Perturbation analysis of the warped disc MHD equations}

In the previous section we found a set of one-dimensional MHD equations for the vertical structure of a magnetised warped disc. We proceed by linearising these equations with respect to the warp amplitude $\psi$. The solutions found by this method will be valid only for small warps ($\psi \ll 1$).

\subsubsection{The unwarped solution}

The equilibrium state is chosen to be an unwarped disc such that $B_y = v_x = v_{z'} = 0$ and there is no jet outflow for simplicity \citep{1997MNRAS.288...63O}. This represents an unwarped accretion disc with a poloidal magnetic field and a purely azimuthal velocity.  \\

The unwarped disc is described by three variables $\lbrace \rho_{0}, B_{x0},v_{y0} \rbrace$ which satisfy the following set of equations:

\begin{equation}
2  v_{y0} + \frac{B_{z'}^2}{\rho_0 }  d _{z '} B_{x0} = 0, 
\end{equation}

\begin{equation}
 z' + \frac{1}{\rho_0} d_{z '} \rho_0 + \frac{B_{z'}^2}{\rho_0} B_{x0} d_{z'} B_{x0}=0,
\end{equation}

\begin{equation}
d_{z '} v_{y0} - q B_{x0} =0.
\end{equation}
The following three associated boundary conditions must also be satisfied:

\begin{equation}
B_{x0} \vert _{z' \rightarrow  \infty} =  \tan i, \label{bc01}
\end{equation}

\begin{equation}
B_{x0} (0) = 0,
\end{equation}
and 

\begin{equation}
\int _{- \infty} ^\infty \rho_0 (z') dz'= 1.
\end{equation} 

The first condition states that the inclination angle $i$ of the poloidal magnetic field has to be specified at large z'.  We only consider $i< \SI{30}{\degree}$, as an inclination angle in excess of this value produces a jet outflow \citep{1982MNRAS.199..883B}. The second condition is a result of the odd symmetry of $B_{x0}$ in $z'$, and the final condition normalizes the surface density of the disc.

\subsubsection{The MHD equations to first order}

 Through inspection of equations \eqref{1dnonlin1}--\eqref{1dnonlin2} and comparison with the hydrodynamic solutions found by \cite{2013MNRAS.433.2403O}, all first-order physical quantities can be expected to vary sinusoidally throughout the orbit. More specifically, the following time-dependence is consistent with the MHD equations \eqref{1dnonlin1}--\eqref{1dnonlin2} to first order in warp amplitude $\psi$: $\lbrace v_{x1} \sim \sin(t'), v_{y1} \sim \cos(t'), B_{x1} \sim \cos(t'), B_{y1} \sim \sin(t'), \rho_{1} \sim \cos(t'), v_{z1'} \sim \sin(t') \rbrace$.\\
 
We now assume the above time-dependence for all six variables and remove the relevant factors of $\sin(t')$ or $\cos(t')$ from the linear equations. Eliminating the time-dependence in this way, the first-order MHD equations reduce to a set of six interrelated first-order ordinary differential equations for the vertical structure of the disc:

\begin{equation}
 v_{x1} - 2  v_{y1} = - \frac{1}{\rho_0} d_{z'} \rho_0 + \frac{B_{z'}^{2}}{\rho_0} \left(  d_{z '} B_{x1} -  \frac{\rho_1}{\rho_0} d_{z '} B_{x0}   -  B_{x0} d_{z'} B_{x0}\right), \label{eqlinstart}
\end{equation}

\begin{equation}
-v_{y1} = -(2-q) v_{x1} + \frac{B_{z'} ^{2}}{\rho_0}d_{z'}  B_{y1} -v_{z1'} d_{z'} v_{y0},
\end{equation}

 \begin{equation}
 \begin{split}
 v_{z1'} =    2  v_{y0} - \frac{1}{\rho_0} d_{z'} \rho_1 + \frac{\rho_1}{\rho_0 ^2} d_{z'} \rho_0  + 
 \frac{B_{z'} ^{2}}{\rho_0 } \biggl( -  B_{x0} d_{z'} B_{x1}  \\
 - B_{x1} d_{z'} B_{x0} +  B_{x0} d_{z'} B_{x0} \frac{\rho_1}{\rho_0} +  d_{z'} B_{x0} \biggr) ,
\end{split} 
 \end{equation}
 
 \begin{equation}
- \rho_1 +  d_{z'} (\rho_0 v_{z1'})= 0,
\end{equation} 

\begin{equation}
- B_{x1} =   d_{z'} v_{x1}
- v_{z1'} d_{z'} B_{x0}
 - B_{x0}  d_{z'} v_{z1'},
\end{equation}

\begin{equation}
B_{y1} =  d_{z'} v_{y1} - q  B_{x1}. \label{eqlinend}
\end{equation}

The symmetry properties of the first-order problem can be found by the following considerations. It was shown in section 3.1.2 that all physical quantities have a symmetry under the transformation $\lbrace t' \rightarrow t' + \pi, z' \rightarrow -z' \rbrace$. Given that the zeroth-order terms have no time-dependence and the first-order terms have a sinusoidal dependence, one can deduce whether each variable is an even or odd function of $z'$. For example $\rho_1(-z',t'+\pi) = \rho_1(z',t')$ from the point symmetry of the warped disc, but the sinusoidal time-dependence of $\rho_1$ implies that $\rho_1(z',t'+\pi) = -\rho_1(z',t')$. We may thus conclude that $\rho_1(-z',t') = -\rho_1(z',t')$, or $\rho_1$ is an odd function of $z'$. In summary, $\lbrace \rho_0,v_{y0},v_{z1'},B_{x1},B_{y1} \rbrace$ are even functions of $z'$, while  $\lbrace B_{x0},\rho_1,v_{x1},v_{y1} \rbrace$ are odd functions of $z'$. These symmetry considerations lead to boundary conditions \eqref{bcstart}--\eqref{bcend}.\\

Equations \eqref{eqlinstart}--\eqref{eqlinend} specify six first-order differential equations with one free dimensionless parameter, $B_{z'}$, which fixes the strength of the magnetic field. This is the same dimensionless parameter used in the study of jet launching from isothermal discs by \cite{2012MNRAS.423.1318O}. In addition we require six boundary conditions in order to construct a numerical solution for the vertical structure of the disc. Symmetry considerations imply that the equations need only be solved above the midplane and immediately fix three boundary conditions at the midplane. The magnetic field at the upper surface must be prescribed, leading to a further two boundary conditions \eqref{bcstart}--\eqref{bc2}.\\

  We chose to impose a rigid lid on the upper disc surface corresponding to the final boundary condition \eqref{bcend}. An analysis of the far-field solution of equations \eqref{eqlinstart}--\eqref{eqlinend} implied that the high-density region of the disc is insensitive to any reasonable boundary condition on the disc outflow at sufficiently large $z'$. Numerical tests (see section 5.1) confirmed this result, validating our choice of boundary condition.\\

In summary the boundary conditions on the problem are:

\begin{equation}
B_{x1}(\infty) = A, \label{bcstart}
\end{equation}

\begin{equation}
B_{y1}(\infty) = B, \label{bc2}
\end{equation}

\begin{equation}
\rho_{1}(0) = 0,
\end{equation}

\begin{equation}
v_{x1}(0) = 0,
\end{equation}

\begin{equation}
v_{y1}(0) = 0,
\end{equation}

\begin{equation}
v_{z1}(\infty) = 0 , \label{bcend}
\end{equation}
where $A$ and $B$ are the limiting values of the radial and toroidal field far above the disc mid-plane. In the context of this model, $A$ and $B$ are free parameters. Physically, $A$ and $B$ need to be determined by solving for the global magnetic field in the force-free region far above and below the disc and solving for the disc's internal structure in a self-consistent manner. Such a problem is global in character and hence beyond the scope of this work, although we will discuss it in a forthcoming publication concerned with bending equilibrium magnetic field configurations.

\subsection{The magnetic field and physical relevance}

Having reduced the MHD equations to a set of linear one-dimensional equations, it is informative to step back and consider the structure of the magnetic field described by this model. Due to the time dependence we have assigned to the magnetic field, the field must be anchored within the disc. Were the magnetic field not anchored within the disc, the differential rotation would inevitably lead to far more complex magnetic field configurations. Beyond being anchored in the disc, we have chosen to model the equilibrium magnetic field as purely poloidal. This assumption is made for mathematical simplicity and also to eliminate the transport of angular momentum by ether an outflow or an exchange of magnetic torques. The extent to which the magnetic field bends is a free parameter of the model. A similar magnetic field structure can be found in \cite{2012MNRAS.424.2097G}\\

If the dimensionless magnetic field strength $B_{z'} \lesssim 1$ we are no longer justified in neglecting the MRI. While a more detailed discussion of the MRI is given in section 4.3, we  briefly mention that due to the neglect of the MRI, this model is of greatest relevance at plasma betas of order unity where the MRI is suppressed. Magnetic fields of this strength are often invoked to explain observed jet emissions (For example, see \cite{2012A&A...548A..76M},\cite{2012MNRAS.423.1318O},\cite{2013ApJ...765..149C}, and \cite{2010MNRAS.401..177C}). \\

Recent studies of protostellar disc formation from molecular cloud core collapse by \cite{2015MNRAS.451..288L}, \cite{2017MNRAS.467.3324L} indicate that protostellar discs may form with plasma betas of order unity. In addition, the continued study of magnetically arrested discs \citep{2003PASJ...55L..69N} around black holes further motivate the study of strongly magnetised warped disc dynamics. It is in this strong regime that the results of this paper are most applicable.

\section{The special case of the purely vertical magnetic field}

\subsection{Solution via a spectral method}

	The simplest magnetised warped disc model is one in which the inclination angle $\tan i$ of the magnetic field at the disc boundary is zero, implying the magnetic field above and below the disc is perpendicular to the disc surface. This corresponds to setting boundary conditions \eqref{bc01}, \eqref{bcstart} and \eqref{bc2} to zero. This model is relatively easy to analyse, yet demonstrates many of the qualitative differences between purely hydrodynamic and magnetised warped discs.\\

Under these conditions, the unwarped density profile is given by

\begin{equation}
\rho_{0}(z') = \frac{1}{\sqrt{2\pi}} e^{-z'^{2}/2}
\end{equation}
and the first-order equations reduce to

\begin{equation}
\frac{d^2 v_{x1}}{d z'^2} = -\frac{\rho_0}{B_{z'} ^2}(v_{x1}-2v_{y1} - z'), \label{coup1}
\end{equation}

\begin{equation}
\frac{d^2 v_{y1}}{d z'^2} = \frac{\rho_0}{B_{z'} ^2}(2v_{x1}-(1+2q)v_{y1}-qz').\label{coup2}
\end{equation}

The coupled pair of differential equations \eqref{coup1} and \eqref{coup2} were solved via a spectral method as outlined below.\\

 Using the set of basis functions $\lbrace y_n(z') = P_{2n-1} (\tanh z')\rbrace$, the first twelve eigenvalues and eigenfunctions of the following Sturm-Liouville problem were found via a combined Rayleigh-Ritz and Gram-Schmidt procedure:

\begin{equation}
\frac{d^2 u}{d z'^2} +\mu \rho_0 u = 0, \label{basis}
\end{equation}

where we impose the boundary conditions $u(0) = 0$ and $\frac{dv}{dz'}=0$ at large $z'$. Let this eigenfunction basis be $\lbrace u_{i} \rbrace$, with associated eigenvalues $\mu_i =\lbrace 3.363, 22.30, 57.12, 107.8, \dots \rbrace$. The eigenfunctions $u_i$ are plotted in Figure 3 (see also \cite{2010MNRAS.406..848L}).\\

 Equations (43) and (44) together with a vanishingly small magnetic field perturbation at large $z'$ imply that velocity perturbations $v_{x1}$ and $v_{y1}$ must have a vanishingly small derivative at the upper boundary. Through symmetry arguments one can deduce that $v_{x1}$ and $v_{y1}$ are odd in $z'$ and so vanish at $z'=0$. Therefore, $v_{x1}$ and $v_{y1}$ satisfy the same boundary conditions as the eigenfunctions $u_i$. Let us project the velocity perturbations $v_{x1}$, $v_{y1}$ and the function $f(z') = z'$ onto this basis in the following way:\\

\begin{figure}
  \centering
    \includegraphics[width=0.4\textwidth]{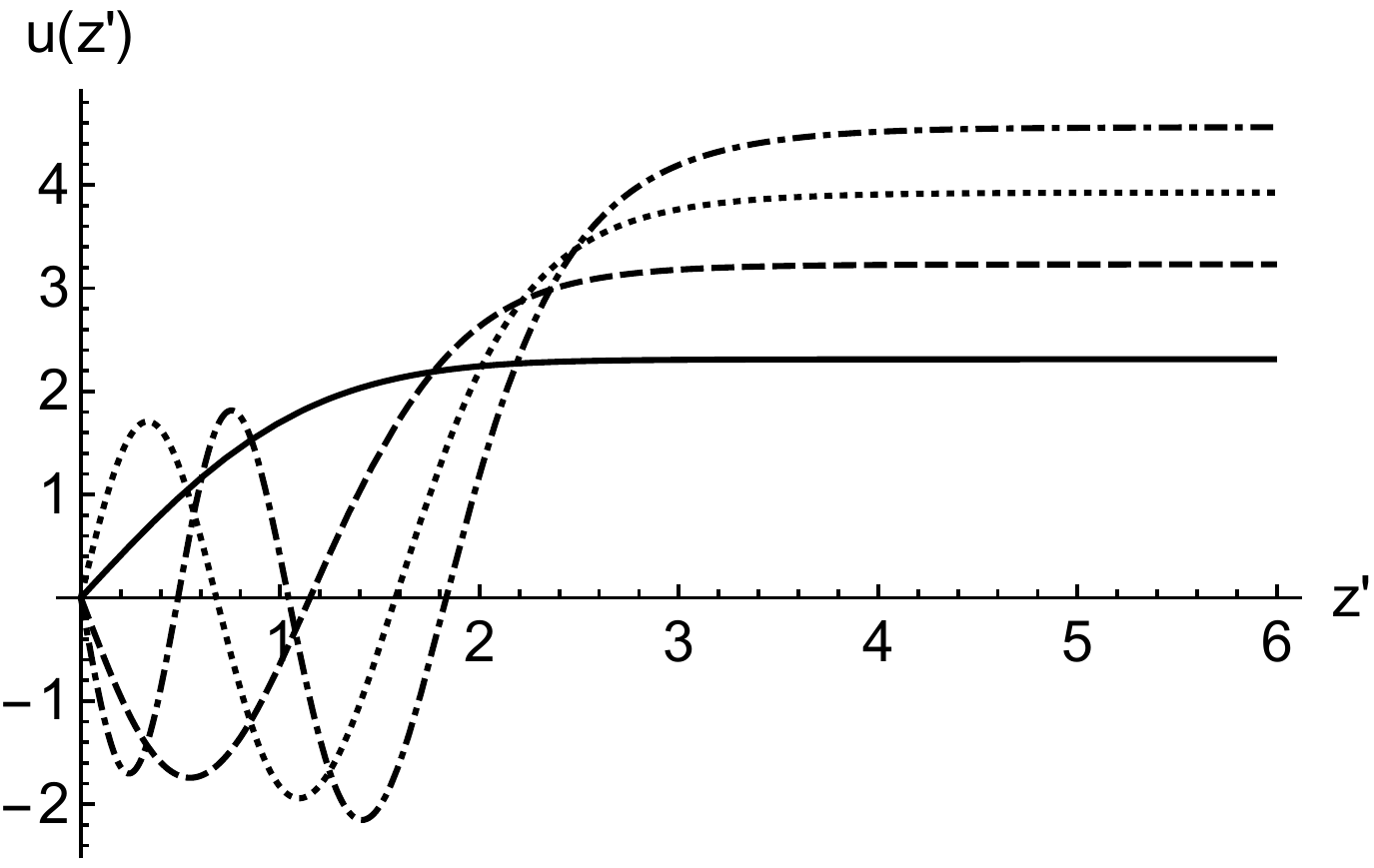}
     \caption{The first four eigenfunctions of equation (71), $u_{1}$,$u_{2}$,$u_{3}$,$u_{4}$. The velocities of the Alfv\'{e}nic-epicyclic modes are proportional to these eigenfunctions.}
\end{figure}
\begin{table*}
  \centering
  \begin{tabular}{c@{\quad}cc}
    & Velocity Perturbation & Magnetic Field Perturbation\\
     $B_{z'}=1.25$ & \includegraphics[width=0.41\textwidth]{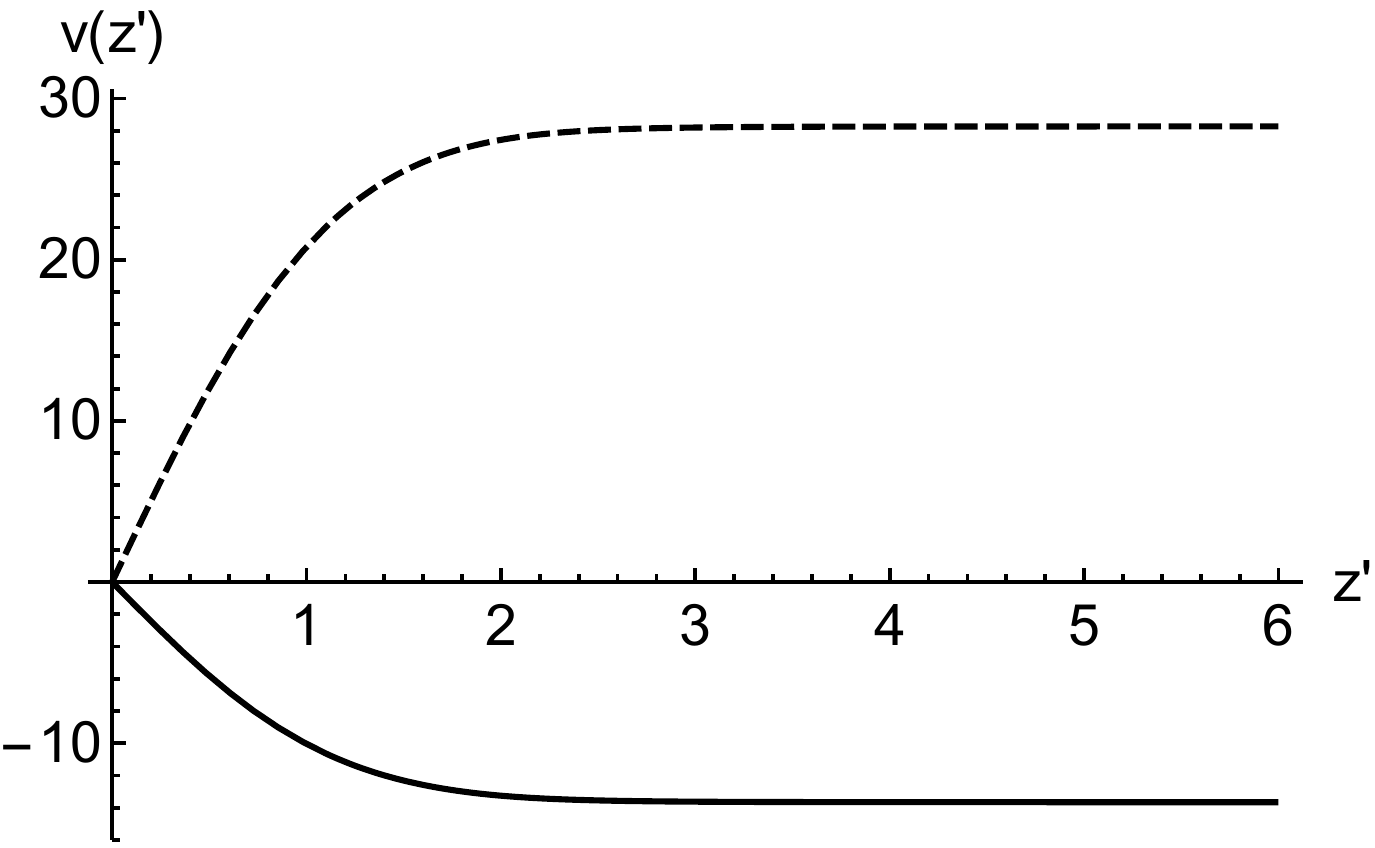}
      & \includegraphics[width=0.45\textwidth]{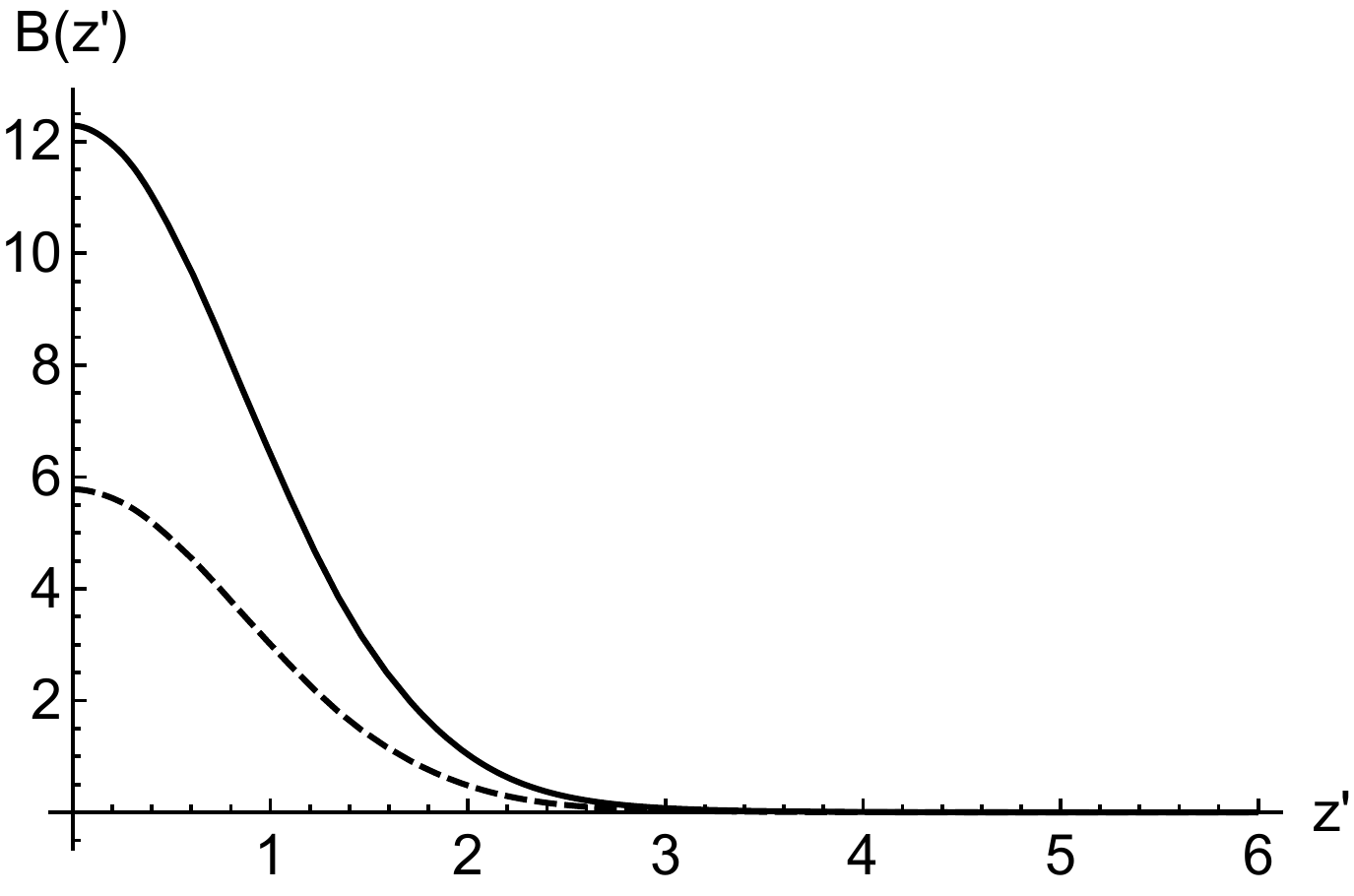}\\ \\
     $B_{z'}=0.12$ & \includegraphics[width=0.41\textwidth]{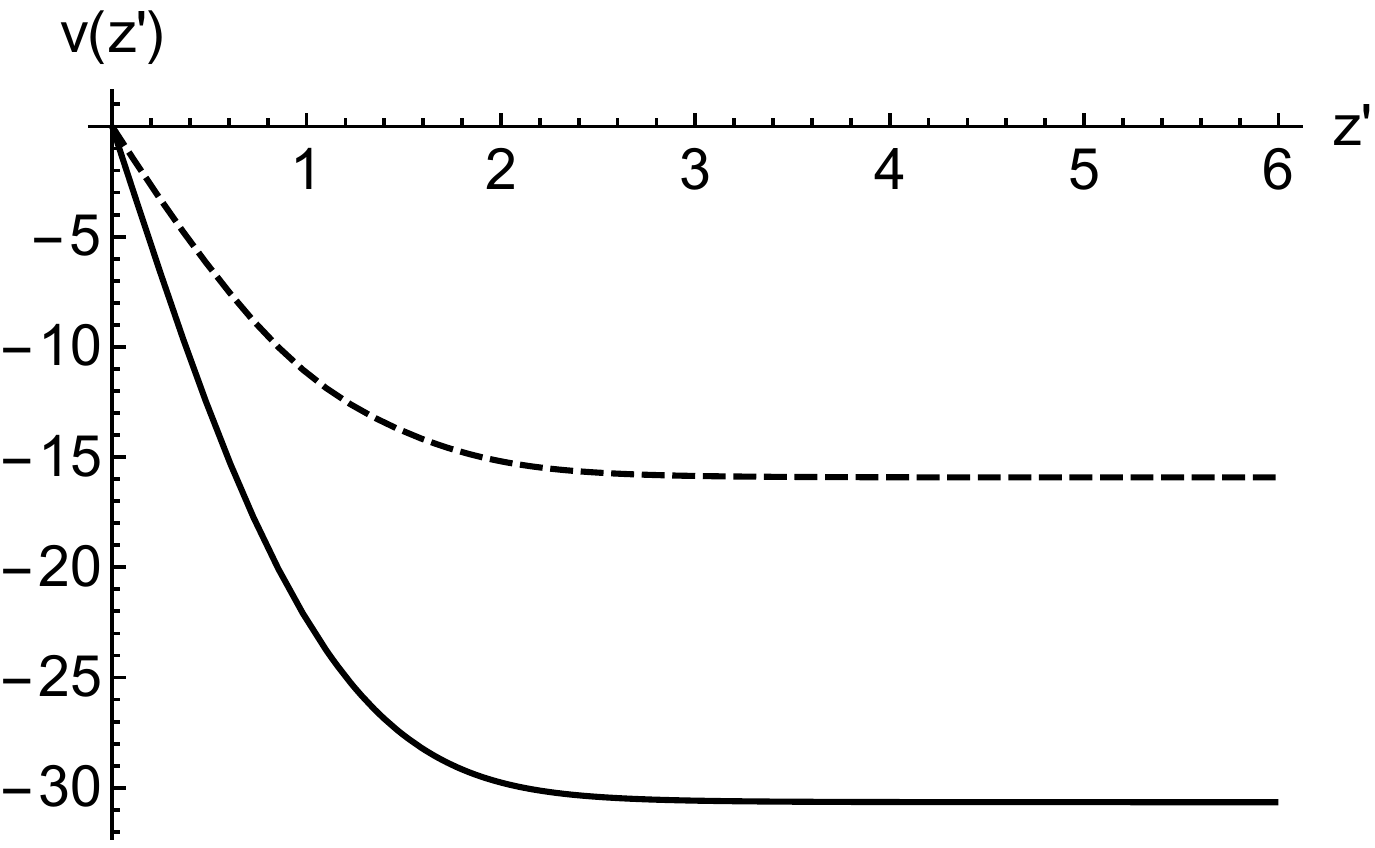} 
      & \includegraphics[width=0.45\textwidth]{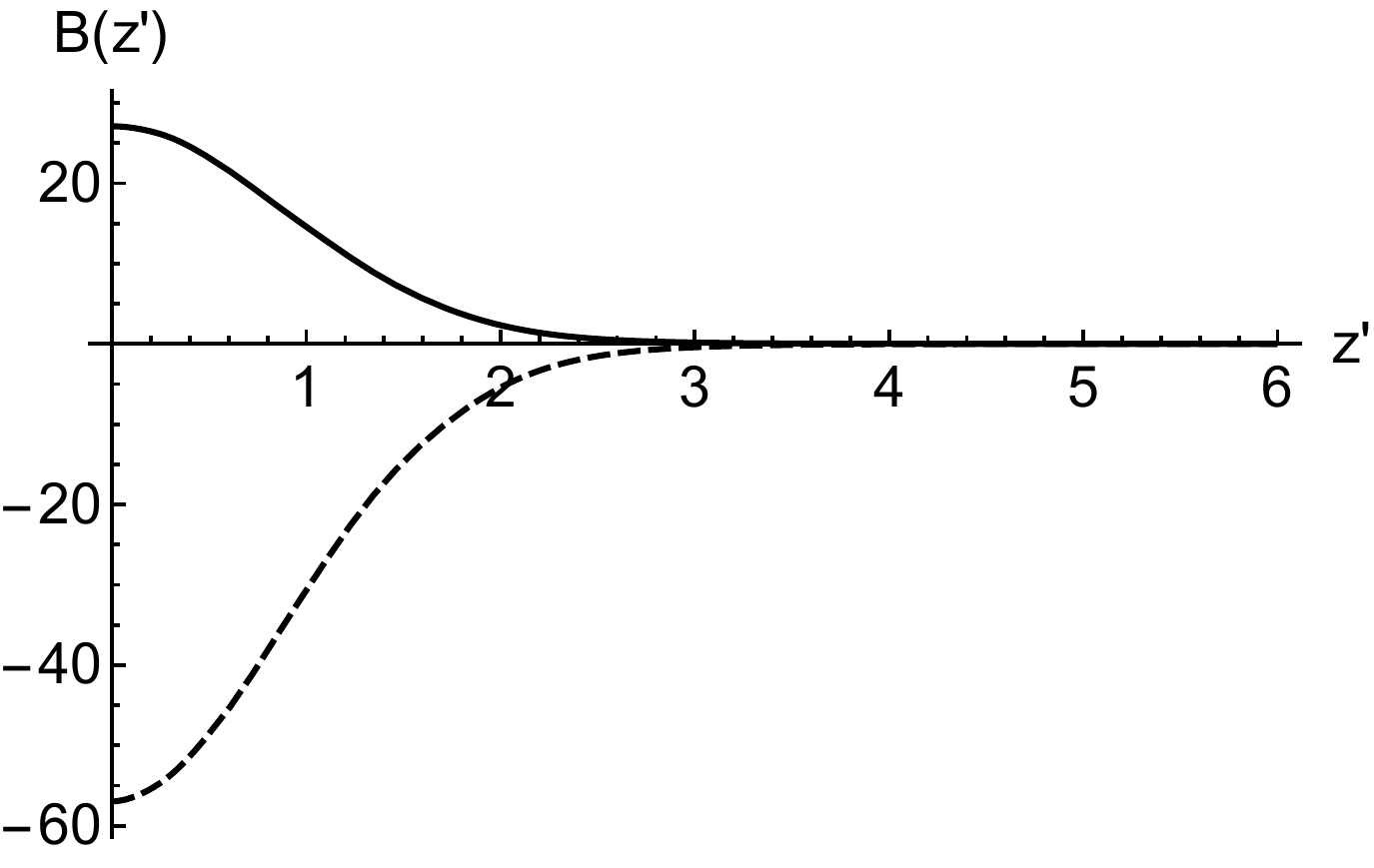} \\ \\
    $B_{z'}=0.45$ & \includegraphics[width=0.41\textwidth]{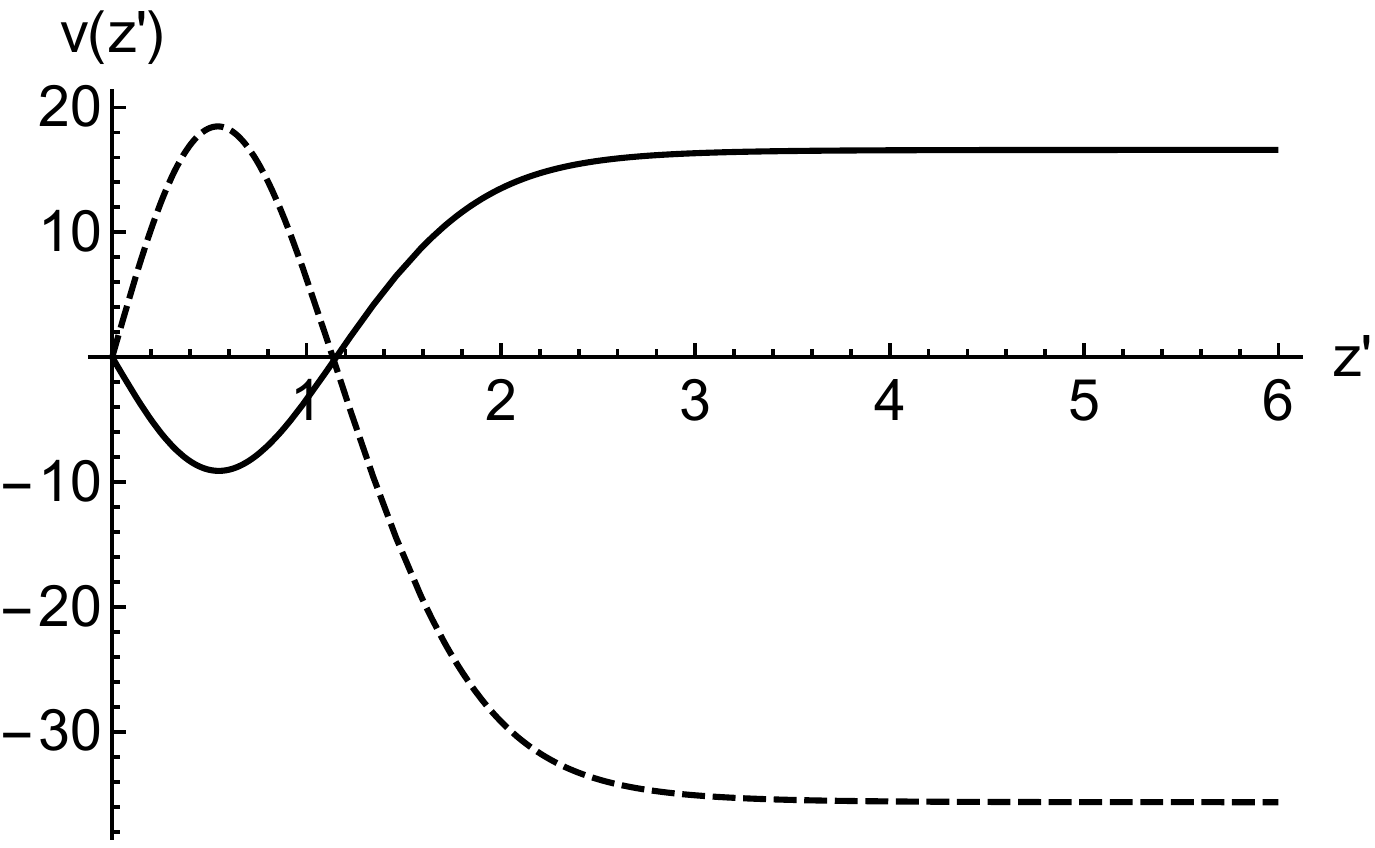}
      & \includegraphics[width=0.45\textwidth]{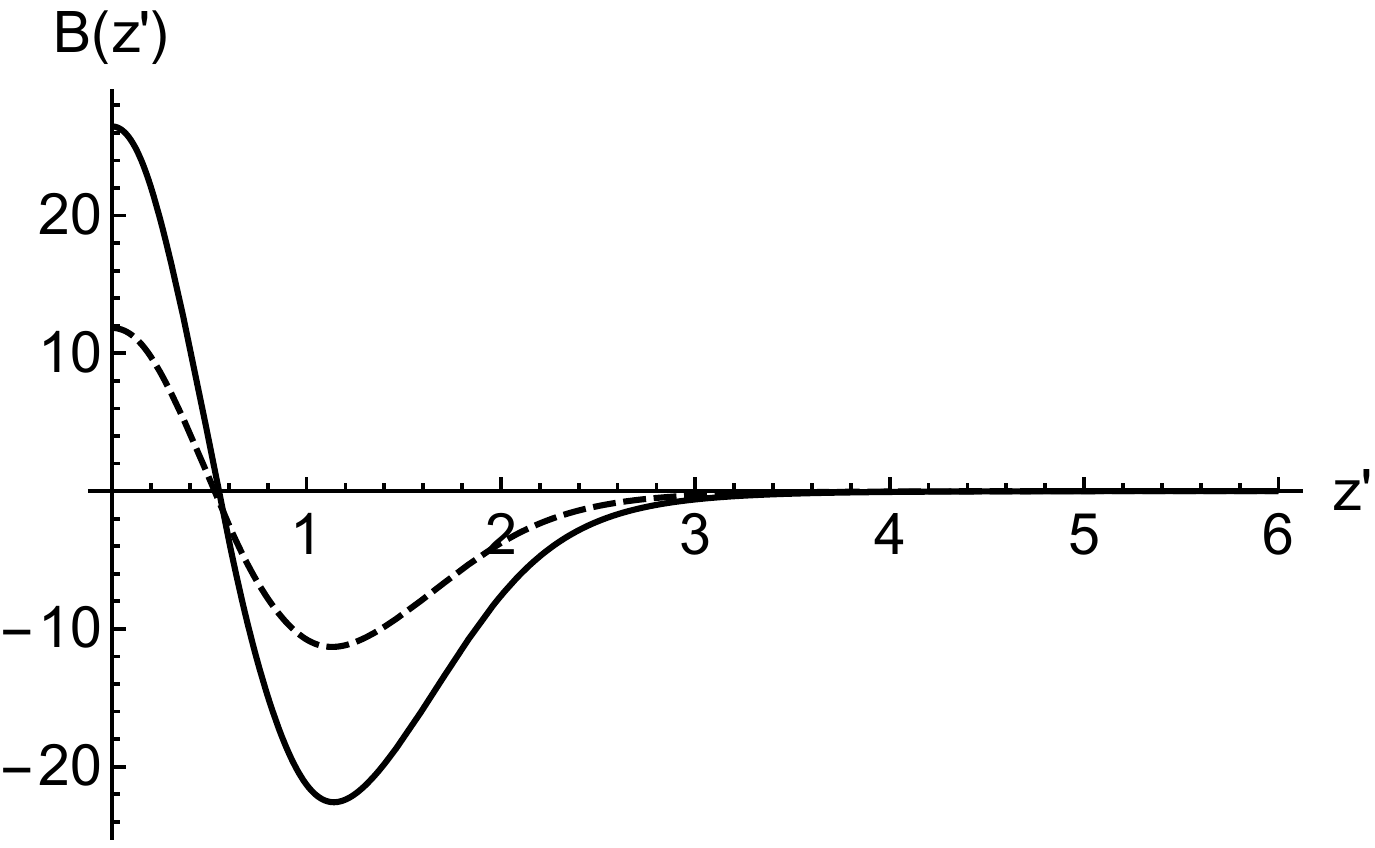}\\ \\
   $B_{z'}=0.04$ & \includegraphics[width=0.41\textwidth]{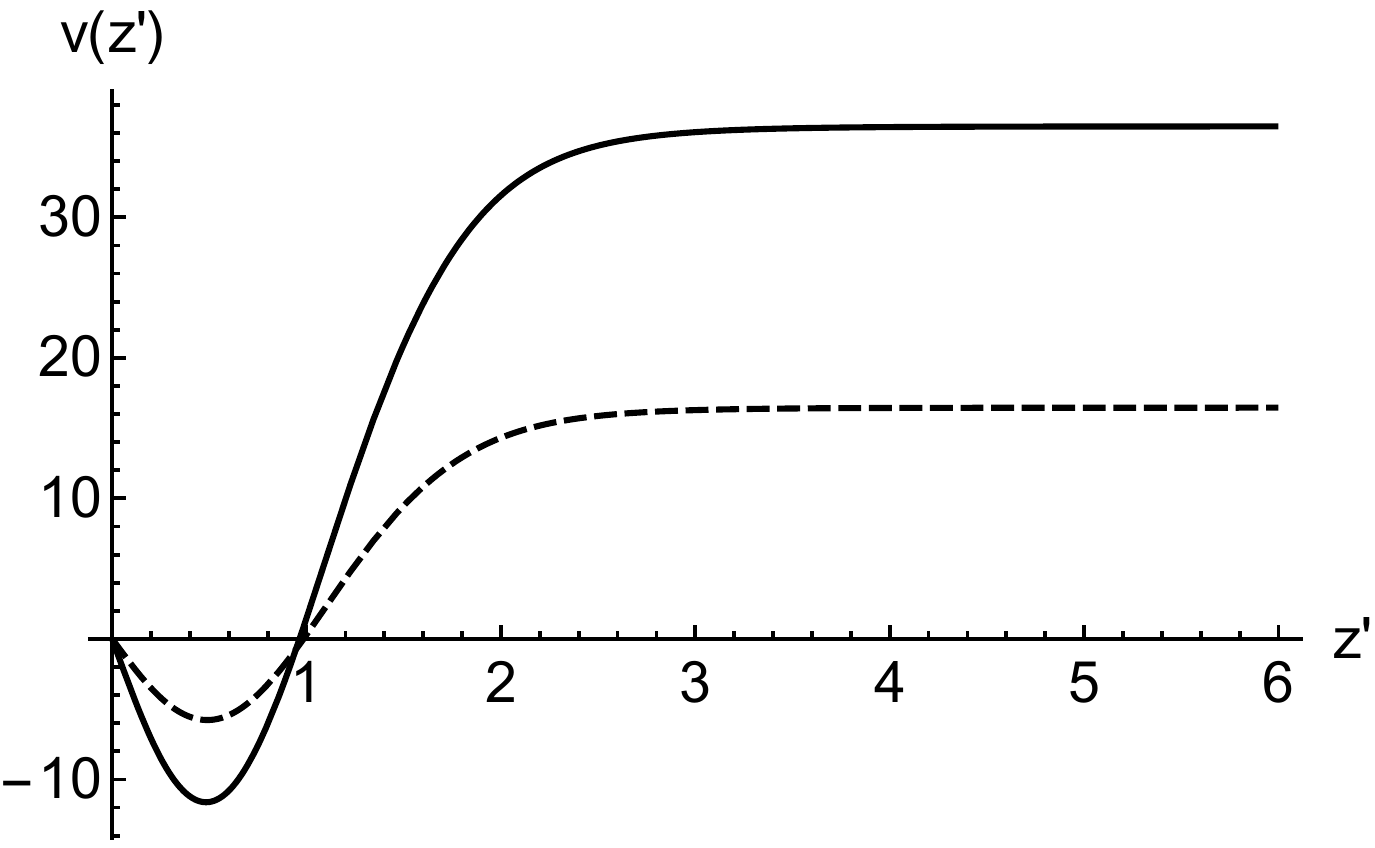}
      & \includegraphics[width=0.45\textwidth]{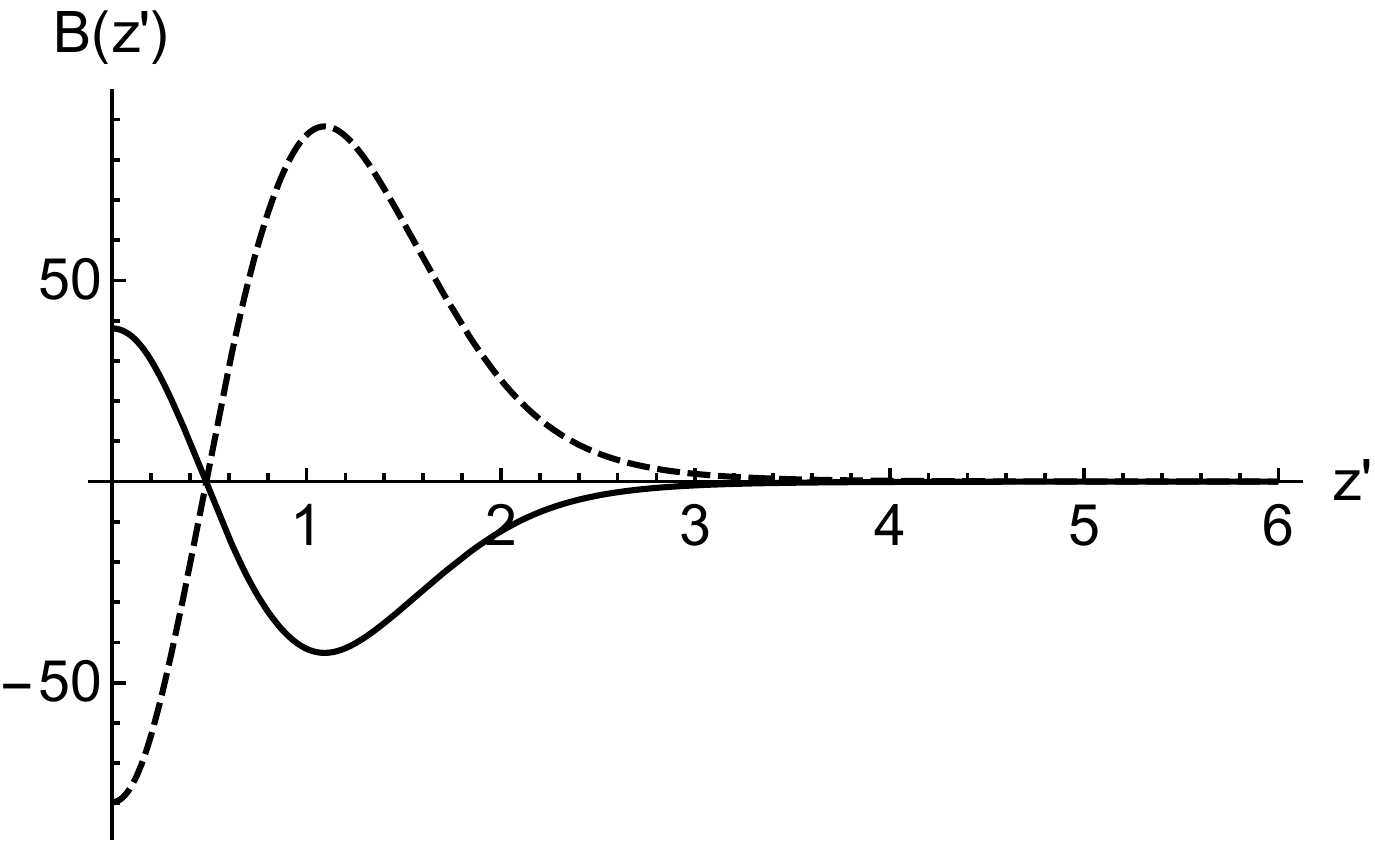}
  \end{tabular}
  \caption{The vertical structure of the first two Alfv\'{e}nic-epicyclic fast and slow modes. The radial and toroidal velocity perturbations $v_{x1}$ and $v_{y1}$ are represented in the left hand column by the solid and dashed lines respectively. The magnetic field perturbations $B_{x1}$ and $B_{y1}$ are represented in the right hand column by the solid and dashed lines respectively. In all cases $q=1.6$ and the magnetic field is varied to be close to the resonant points of the first two Alfv\'{e}nic-epicyclic fast and slow modes. Referencing from the top row down, the resonances shown correspond to the first slow mode, the first fast mode, the second slow mode, and the second fast mode. The associated magnetic field strengths are shown on the left of each row. }\label{figtab}
\end{table*}

\begin{equation}
v_{x1} (z') = \sum _i a_i u_i (z'),
\end{equation}

\begin{equation}
v_{y1}(z') =\sum _i b_i u_i (z'),
\end{equation}

\begin{equation}
z' =\sum _i c_i u_i (z').
\end{equation}

Projecting the function $z'$ onto the eigenfunction basis $\lbrace u_{i} \rbrace$ requires some consideration, as this function does not satisfy the appropriate boundary conditions far from the midplane. However, beyond a few scale heights the density becomes vanishingly small and  the error in approximating $z'$ via the set of eigenfunctions $\lbrace u_i \rbrace$ is negligible.\\

If we let $\lambda_i = \mu_iB_{z'}^{2}$, the MHD equations \eqref{coup1}--\eqref{coup2} give

\begin{equation}
a_i = -\frac{\lambda_i-1}{\lambda_i^2 -2(1+q)\lambda_i + (2q-3)} c_i, \label{ai}
\end{equation}

\begin{equation}
b_i = \frac{\lambda_{i} q + 2-q}{\lambda_i^2 -2(1+q)\lambda_i + (2q-3)} c_i,
\end{equation}
where

\begin{equation}
c_i = \int_0 ^{\infty} z' \rho_0(z') u_{i}(z') dz'. \label{ci}
\end{equation}

The critical aspect of this set of equations is the dependence of the denominator on the magnetic field strength and the shear rate $q$ associated with the disc. Resonances will occur whenever either of the following conditions holds for any eigenvalue $\mu$:

\begin{equation}
B_{z'}^2 \mu = 1+q+\sqrt{4+q^2}, \label{rescond1}
\end{equation}

\begin{equation}
B_{z'}^2 \mu = 1+q-\sqrt{4+q^2}. \label{rescond2}
\end{equation}

These resonances result in arbitrarily large velocities, magnetic fields and internal torques acting within the disc. More generally equations \eqref{ai}--\eqref{ci} imply the existence of a set of modes, henceforth called Alfv\'{e}nic-epicyclic modes, in this magnetised warped disc. They are the result of a coupling between the magnetic field and the epicyclic modes.  Each of these modes is associated with its own distinct vertical structure and a natural frequency given by

\begin{equation}
\omega_{i(fast)}^2 = \lambda_{i} + 2 -q + \sqrt{4\lambda_i + (2-q)^2}
\end{equation}
and 
\begin{equation}
\omega_{i(slow)}^2 = \lambda_{i} + 2 -q - \sqrt{4\lambda_i + (2-q)^2}.
\end{equation}

Whenever there is a coincidence between the natural frequency of a particular Alfv\'{e}nic-epicyclic mode and the orbital frequency ($\omega=1$ in scaled units), a resonance will occur and that mode will dominate the solution.\\

The Alfv\'{e}nic-epicyclic mode velocities are proportional to their respective eigenfunction $u_{i}$, while the magnetic field strengths associated with each mode are proportional to the derivatives of the respective eigenfunction $\frac{d u_{i}}{dz'}$. The radial and azimuthal velocities for the first and second Alfv\'{e}nic-epicyclic modes are shown in Table 1. The absolute magnitude of these velocities becomes arbitrarily high as a resonance is approached, leading to a breakdown of the linear formalism outlined here. An investigation into the fully non-linear theory would be required to explore the behaviour near resonance.  \\

In the Keplerian case ($q$=3/2), which is the case of greatest physical interest, equations \eqref{rescond1}--\eqref{rescond2} reduce to:

\begin{equation}
B_{z'}^2 \mu_{i} = 5, 
\end{equation}

\begin{equation}
B_{z'}^2 \mu_{i} = 0,
\end{equation}
implying that the fast modes are all resonant iff $B_{z'}=0$. This corresponds to the hydrodynamic resonance familiar from earlier work. On the other hand, if the strength of the magnetic field is non-zero this resonance will be lifted and, provided the orbital frequency does not coincide with the natural frequency of a slow Alfv\'{e}nic-epicyclic mode, well-behaved solutions are obtained. The magnetic field strengths associated with these resonances of the slow modes are at $B_{z'} = \lbrace 1.22,0.473,0.296,0.215,\dots \rbrace$, the resonances becoming more and more closely spaced as the magnetic field strength approaches zero.

\subsection{The internal torque}

The internal structure of a magnetised disc with an external magnetic field oriented normal to the disc surface was found in the previous section. The torque acting on the disc can now be calculated. As found in \cite{2013MNRAS.433.2403O} , the torque acting within the disc may be written in the following way:

\begin{equation}
\mathbf{G} = -2 \pi \Sigma c_s^2 r^2\left( Q_1 \mathbf{l} + Q_2 r \frac{\partial \mathbf{l}}{\partial r} + Q_3 r \mathbf{l} \wedge \frac{ \partial \mathbf{l}}{\partial r} \right),
\end{equation}
where $Q_1$,$Q_2$, and $Q_3$ are dimensionless coefficients given by

\begin{equation}
-Q_1 \Sigma c_s ^2 = \left\langle \int (\rho v_x v_y - T_{xy}) dz' \right\rangle_{t'},
\end{equation}

\begin{equation}
-Q_4 \psi \Sigma c_s ^2 = \left\langle e^{i t'} \int [\rho v_x (-\Omega z' - i v_z) + i T_{xz} ]dz' \right\rangle_{t'}, \label{Q4}
\end{equation}
where $Q_4 = Q_2 + i Q_3$. The angular brackets denote that the quantities inside the brackets are time-averaged. $Q_1$ is similar to the usual accretion disc torque, while $Q_2$ corresponds to diffusion of the warp and $Q_3$ causes a dispersive wave-like propagation of the warp. \\

Torque coefficients $Q_1$ and $Q_2$ are related to dissipation. This model does not include viscosity or resistivity, implying both of these coefficients will vanish and only $Q_3$ need be considered. We note that this can be also be proven by considering the symmetry properties of the system as discussed in section 3.1.2. \\

Equation \eqref{Q4} suggests that a radial magnetic stress $T_{xz}$ will contribute to $Q_3$; for a vertical equilibrium field configuration however, this radial stress is exactly balanced by a magnetic stress difference above and below the disk surface. The contribution to the torque from magnetic stresses is therefore of order $O(h/r)$ relative to the angular momentum flux, and is assumed to be small. Only the radial flux of angular momentum need be considered, implying that

\begin{equation}
Q_3 = \frac{1}{2 \pi}\int_0 ^{2\pi} \int \rho_0 v_{x1} z' \sin^2 t' dz' dt' + O(\psi^2). \label{Q32}
\end{equation}

 Projecting $v_{x1}$ and $z'$ onto the eigenfunction basis $\lbrace u_{i} \rbrace$ described above in \eqref{basis} and using the orthogonality relationships of the eigenfunction basis, equation \eqref{Q32} simplifies to

\begin{equation}
Q_3 = \sum_{i} \frac{(1- \lambda_i )}{ \lambda_i ^2 - 2(1+q) \lambda_i + (2q-3)} c_{i}^2, \label{Q33}
\end{equation}
where $\lambda_i = \mu_i B_{z'}^2$. In the hydrodynamic limit, recalling that normalization of density implies $\sum c_i ^2 = 1/2$, we find
\begin{equation}
Q_3 = \frac{1}{2(2q-3)} + O(\psi^2, B^2),
\end{equation}
which is consistent with the result found by \cite{2013MNRAS.433.2403O} for a hydrodynamic warped disc. \\

The internal torque coefficient $Q_3$, as can be seen by the denominator of equation \eqref{Q33}, also diverges near an Alfv\'{e}nic-epicyclic resonance. The dependence of $Q_3$ on the magnetic field strength $B_{z'}$ and shear rate $q$ is shown in figures 4--8. The Alfv\'{e}nic-epicyclic resonances as functions of $q$ and $B_{z'}$ are shown in figure 9 for comparison.\\

 It is worthy of note that in the strong-field limit $ B_{z'}^2\gg1$, $Q_3 \approx \frac{-c_1 ^2}{\mu_1 B_{z'}^2}$. At very large magnetic field strengths the increasing stiffness imbued by the magnetic flux dampens the disc response. Consequently the induced internal flows lower in amplitude and warp propagation slows as the magnetic field strength is increased.\\
 	
The weak field limit is far more subtle. Since the model presented in this paper is concerned with mean-field effects, it is reasonable to expect that at weaker field strengths the hydrodynamic limit for $Q_3$ is recovered. It can be seen in Figure 4 that this holds true at all points except those very near to an Alfv\'{e}nic-epicyclic resonance. At these resonances the warp structure and propagation properties may be  quite sensitive to the magnetic field strength and even a relatively weak magnetic field may have significant consequences for the warp evolution. This has special physical relevance in light of the fact that a hydrodynamic Keplerian disc is precisely at one such resonance (or alternatively, the resonance of a Keplerian hydrodynamic disc could be considered as the pile-up of fast Alfv\'{e}nic-epicyclic modes as the magnetic strength falls to zero as shown in Figure 8). Warp propagation in a Keplerian low-viscosity disc may therefore be quite sensitive to the introduction of even a weak magnetic field. In the Keplerian weak-field limit $B_{z'}^2\ll1$, the case of most physical interest, we note that $Q_3 \approx \frac{-c_1 ^2}{5\mu_1 B_{z'}^2}$. \\
 	 
 	  In a more realistic model one might expect the mean-field effects described above to compete with viscous effects including those that might arise from MRI turbulence. In some circumstances, these non-ideal effects may eliminate any distinction between the MHD model presented here and the corresponding hydrodynamic model. Further investigation is required to determine what factors influence warp propagation in a weakly magnetised disk. \\

\begin{figure}
  \centering
    \includegraphics[width=0.4\textwidth]{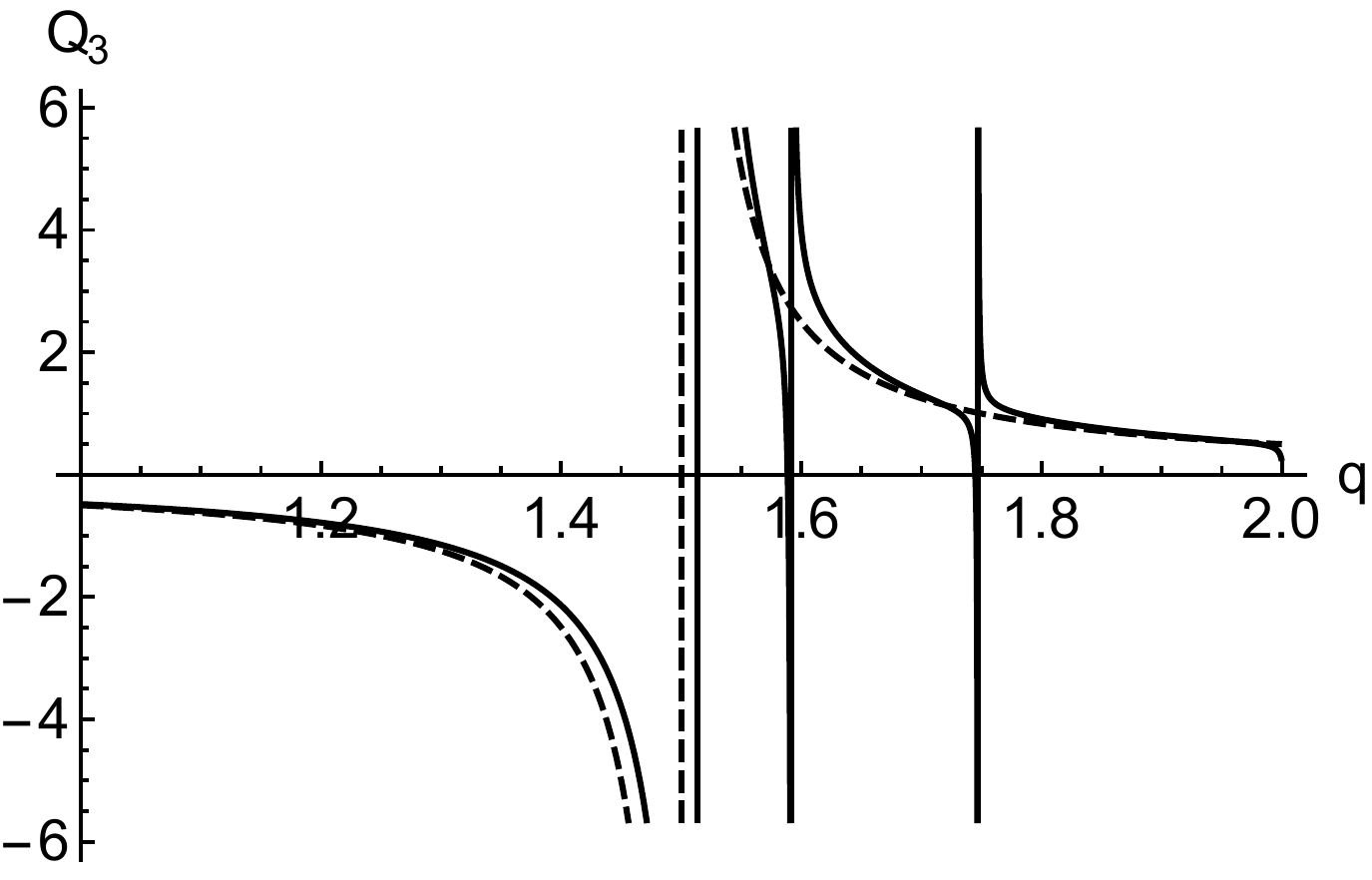}
    \caption{The internal torque coefficient $Q_3$ as a function of $q$ at magnetic field strength $B_{z'}=0.04$. The dotted line shows $Q_3$ for a hydrodynamic disc. The discontinuities at $q>1.5$ correspond to resonances of fast Alfv\'{e}nic-epicyclic modes.}
\end{figure}

\begin{figure}
  \centering
   \includegraphics[width=0.4\textwidth]{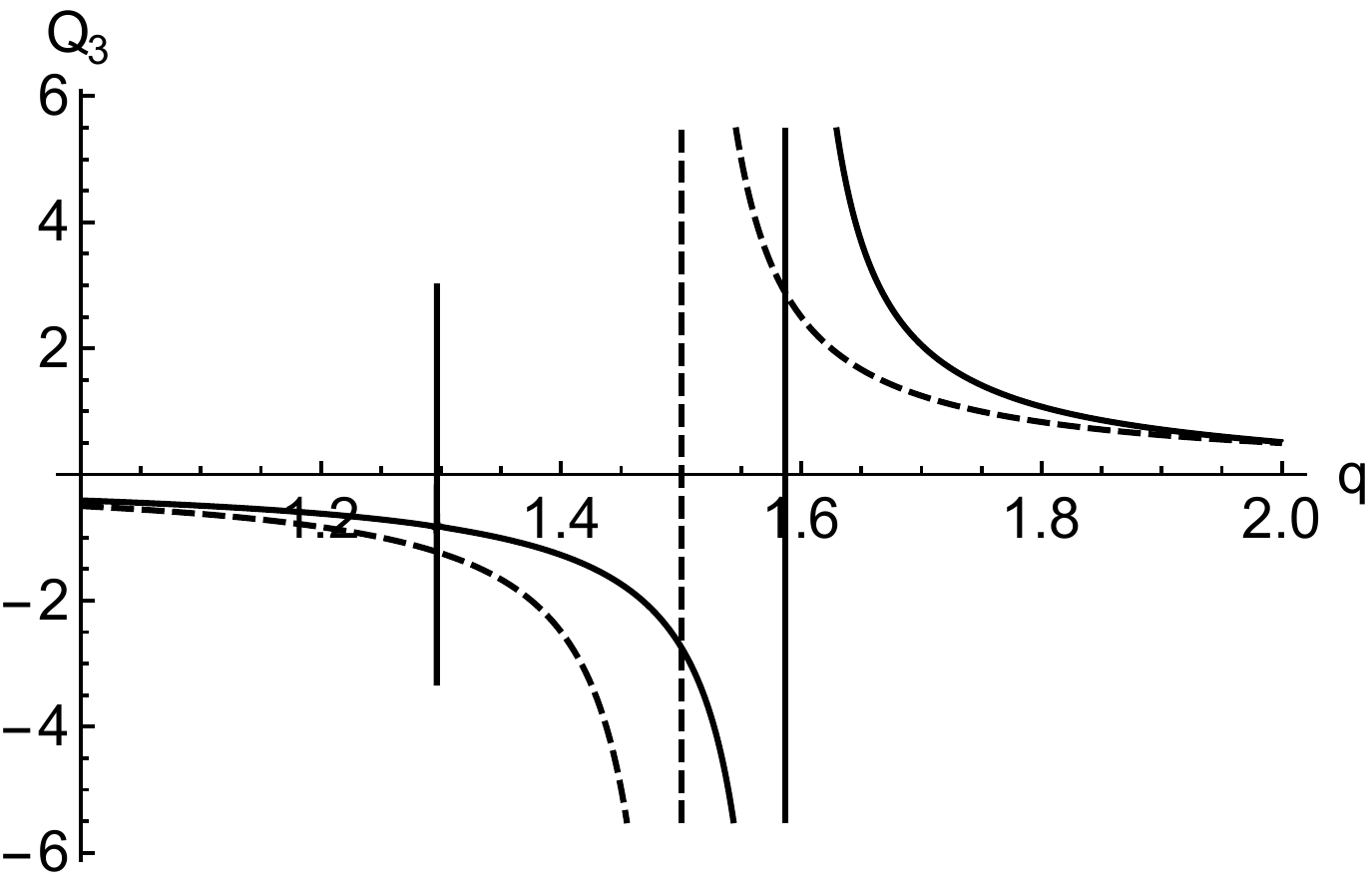}
     \caption{The internal torque coefficient $Q_3$ as a function of $q$ at magnetic field strength $B_{z'}=0.1$. The central discontinuity originally at $q=1.5$ has been displaced rightward, lifting the hydrodynamic resonance at $q$=1.5.}
\end{figure}

\begin{figure}
  \centering
   \includegraphics[width=0.4\textwidth]{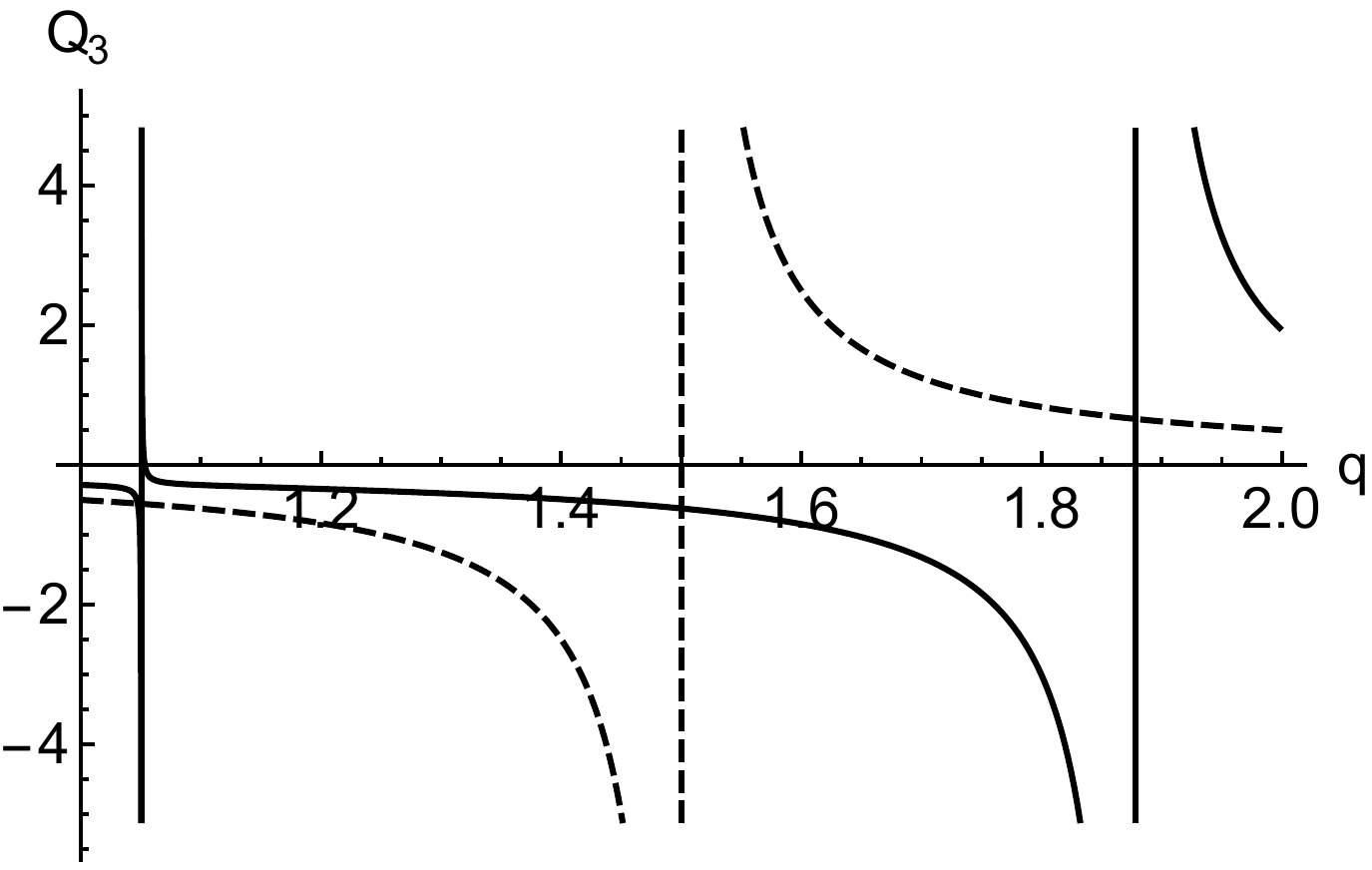}
 \caption{The internal torque coefficient $Q_3$ as a function of $q$ at magnetic field strength $B_{z'}=0.2$. The central discontinuity has moved well past $q=1.5$. Near $q=1.0$ there is a discontinuity associated with a slow Alfv\'{e}nic-epicyclic mode.}
\end{figure}

\begin{figure}
  \centering
    \includegraphics[width=0.4\textwidth]{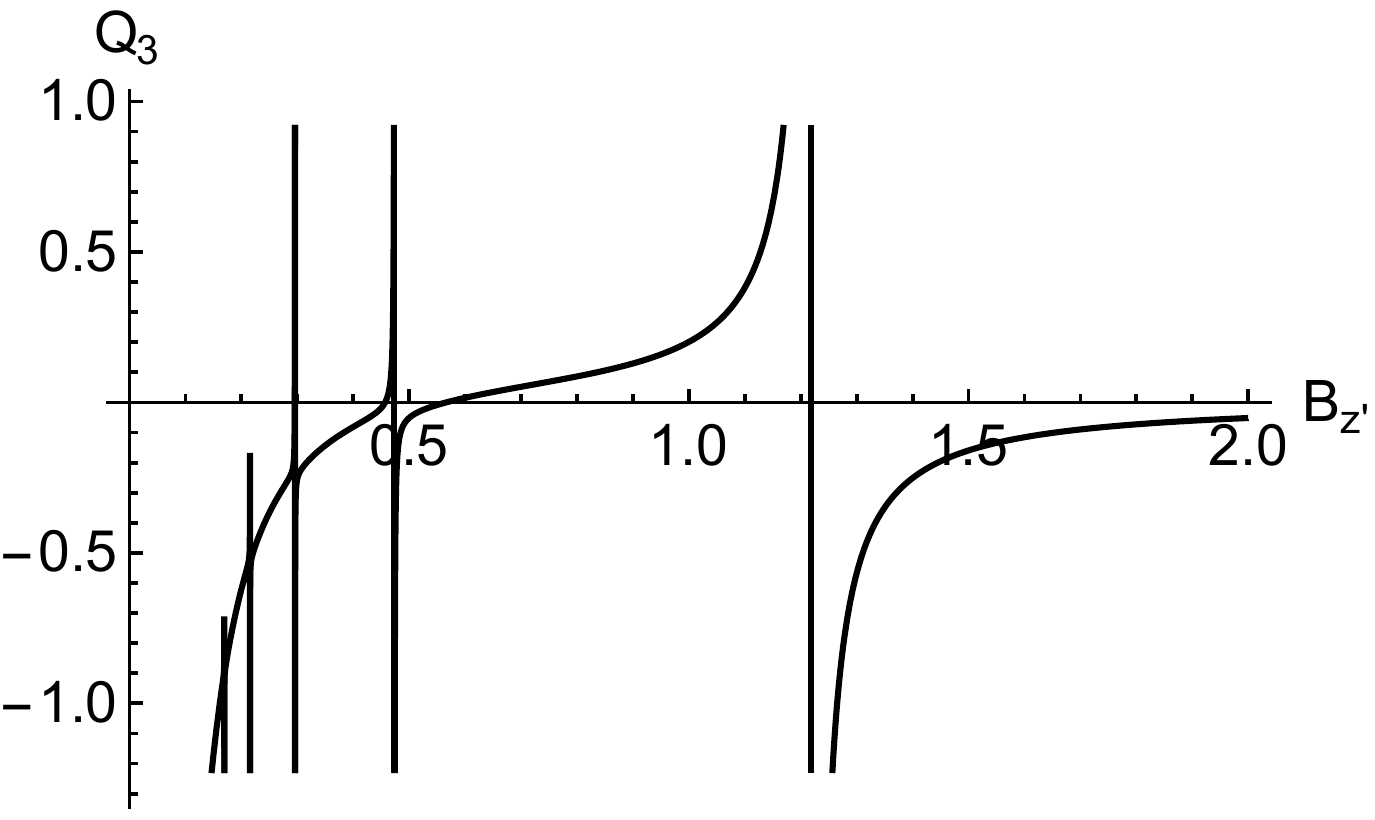}
     \caption{The internal torque coefficient $Q_3$ as a function of magnetic field strength $B_{z'}$ in scaled units for a Keplerian disc. The most significant resonance occurs at $B_{z'}=1.22$, corresponding the ground state Alfv\'{e}nic-epicyclic mode.}
\end{figure}

\begin{figure}
  \centering
    \includegraphics[width=0.50\textwidth]{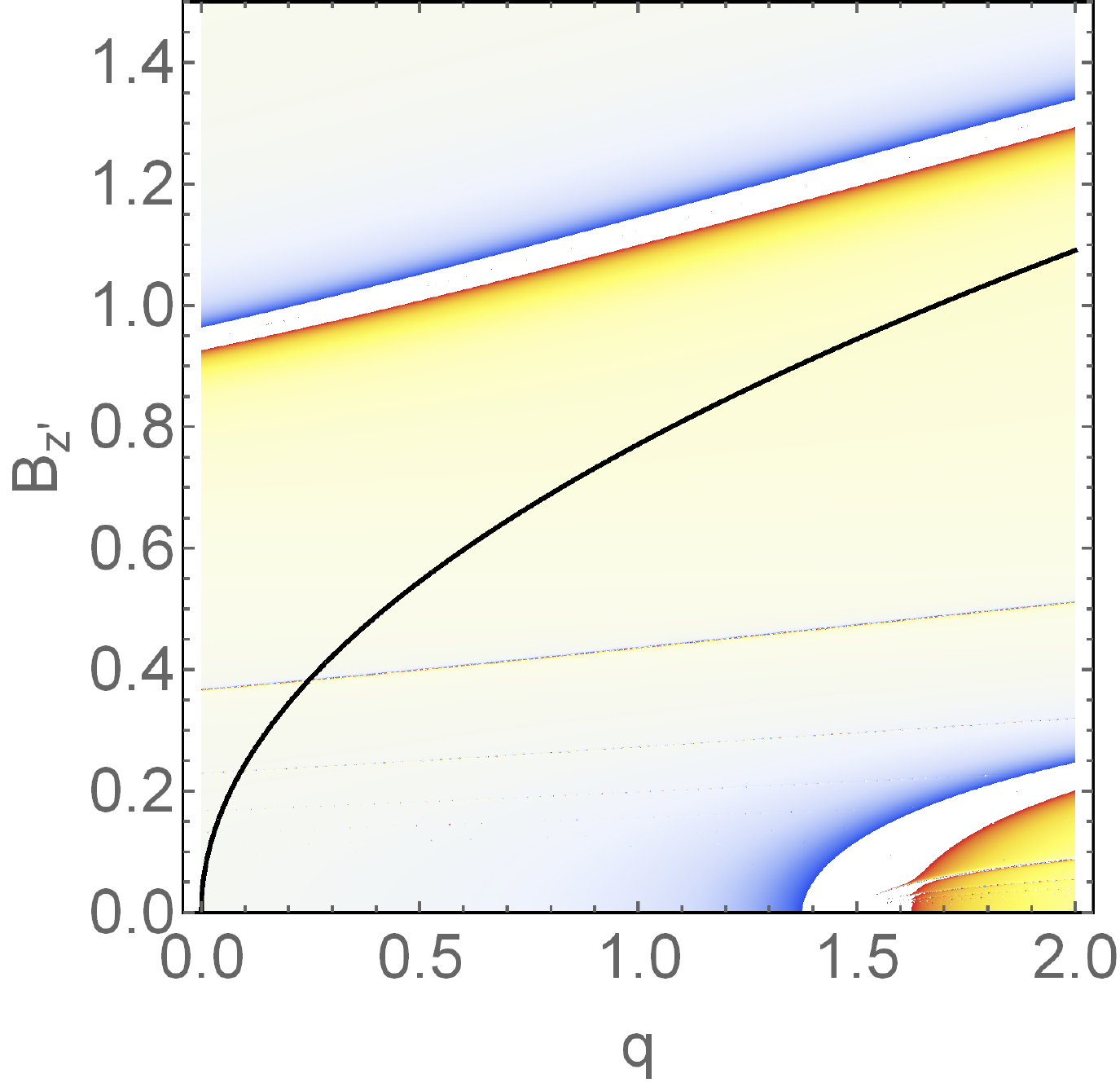}
    \caption{Torque coefficient $Q_3$ as a function of shear rate $q$ and magnetic field strength $B_{z'}$. White regions indicate large torques that fall beyond the plot range. Due to poor resolution, many of the higher order resonances are not clearly visible. Large values of $Q_3$ can be observed near the first slow mode resonance where $B_{z'} \approx 1$ and near the first fast mode resonance. The fast mode resonance is responsible for the anomalous behaviour of Keplerian hydrodynamic warped discs. The MRI stability curve is shown in black.  }
\end{figure}

\begin{figure}
  \centering
    \includegraphics[width=0.4\textwidth]{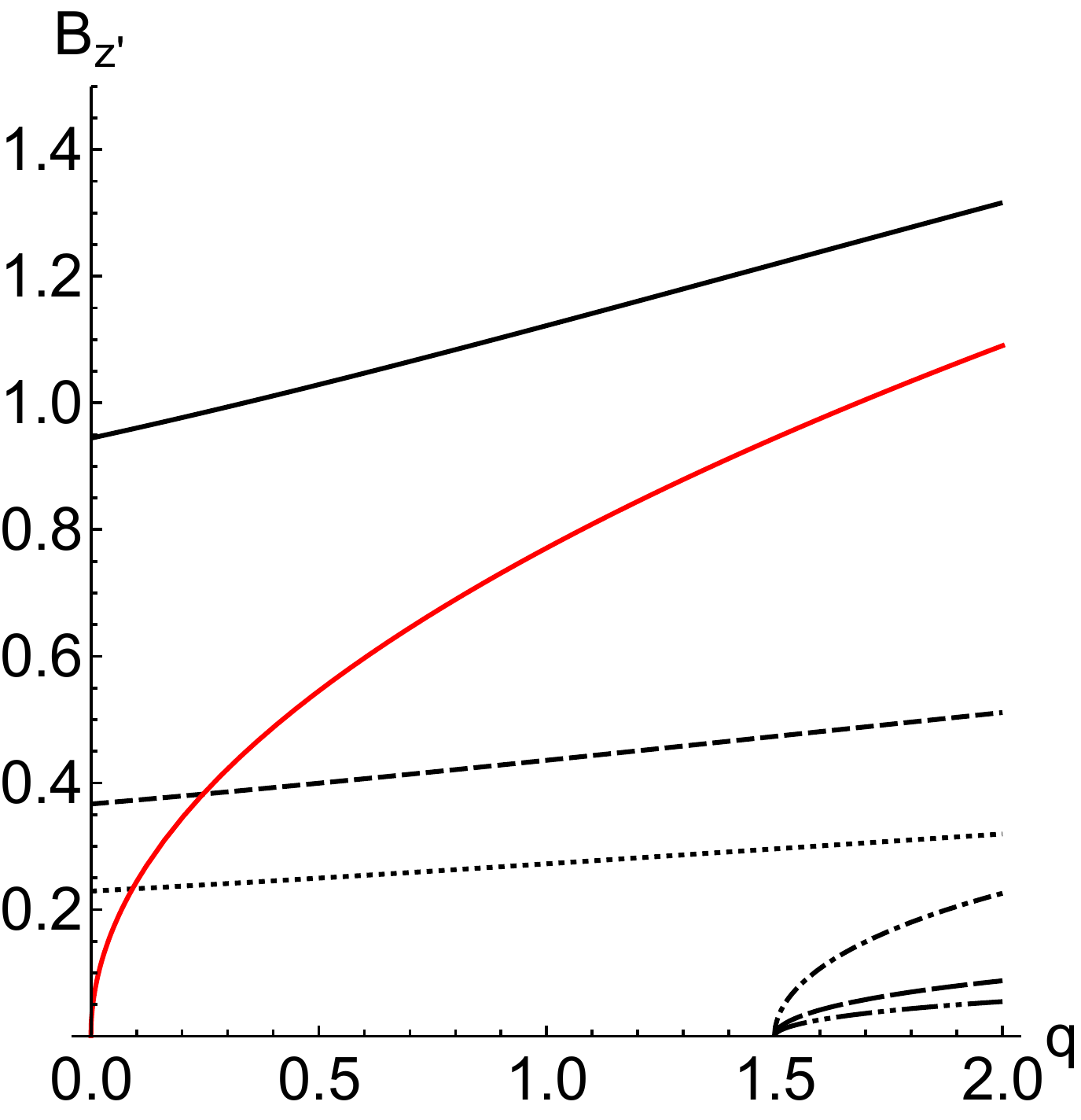}
    \caption{The Alfv\'{e}nic-epicyclic resonances in the $q$-$B_{z'}$ plane. The first three slow resonances are the top three nearly straight lines in this figure. The bottom three curves, with a common meeting point at $q=3/2$ for the hydrodynamic case, are the first three fast Alfv\'{e}nic-epicyclic resonances. The MRI stability curve is shown in red.}
\end{figure}

	Although viscosity has not been included in any of the derivations thus far, let us consider the effect of a magnetic field on the internal torque coefficient $Q_2$. It was found in \cite{2013MNRAS.433.2403O} that the torque coefficient $Q_2$ in a Keplerian disc grows inversely with viscosity as the hydrodynamic resonance is approached. It is expected, subject to further analysis, that the presence of a magnetic field will lift the resonance of a Keplerian hydrodynamic disc and therefore remove the divergence in torque coefficient $Q_2$, reducing the anomalously quick diffusion of the warp in this case.\\
	
	We would like to emphasise for completeness that it is only for the vertical equilibrium field configuration that there is no explicit magnetic contribution to the torque. Warp propagation in discs with a `bending' field ($B_{x0} \neq 0$) differs from the vertical field case in several respects. Notably, the magnetic field can potentially play a greater role in the transport of angular momentum through the disc. We refer the reader to an upcoming paper for details of warp propagation in more general magnetic field configurations. \\

\subsection{Effects of the magnetorotational instability (MRI)}

The MRI in an accretion disc can be expected to exist for a range of magnetic field strengths $B_{min}<B_{z'}<B_{max}$. The upper limit of the MRI-unstable region $B_{max}$ may be found by a standard analysis of the ideal MHD equations as first derived in \cite{1991ApJ...376..214B}. It is informative to note the connection between their analysis and the analysis presented here. While investigations of the MRI are primarily focused on the unstable normal modes of a magnetised accretion disc, the chief focus of this work has been the resonant forcing of stable modes. However, when the natural frequency of a stable Alfv\'{e}nic-epicyclic mode transitions from a real to an imaginary number, that mode becomes unstable corresponding to the onset of the MRI. From equation (64) one can deduce that this transition will occur when

\begin{equation}
\mu_{1} B_{z'}^2 = 2q,
\end{equation}
where $\mu_{1}$ is the smallest of the eigenvalues defined by equation \eqref{basis}. In the case of a vertical magnetic field and a Keplerian disc, we can conclude that

\begin{equation}
B_{max} ^2 = \frac{q}{1.68}.
\end{equation}

Though formulated slightly differently, this corresponds exactly to the result found by \cite{1994MNRAS.270..138G} for isothermal discs\footnote{We are in agreement with the general form for the Alfv\'{e}n velocity at the midplane, $V_{0z}^2 > 2q \Omega_0^2 H^2/E_1$, as well as the lowest order non-zero eigenvalue $E_1 \approx 1.34$. However, we do not agree on the exact stability condition $V_{0z}^2 > 0.45 c_s ^2$ that they quote for $q=3/2$.}. Given that the resonance associated with the lowest order eigenfunction of equation (54) occurs for a Keplerian disc at $B_{z'} \approx 1.22$ while the MRI is expected below $B_{z'} \approx$ 0.94, one can see that a large-scale magnetic field may have a significant effect on the warp dynamics even in the MRI stable region. This is an important result because if it was discovered that the resonances discussed in the previous chapters only existed in the MRI unstable region $B_{z'}<B_{max}$, one would be uncertain as to whether these resonances are likely to exist once the effects of the MRI are taken into account.\\

There are a few shortcomings with the analysis above. Firstly, we found an identical MRI stability condition to that found by \cite{1994MNRAS.270..138G}. This is unsurprising as the MRI has emerged as an instability in our linear perturbation of an unwarped disc state. Non-linear warped disc solutions may not have the same MRI stability conditions and require a non-linear analysis beyond the scope of this work. Secondly, the value of $B_{max}$ found is only valid for the particular vertical magnetic field configuration considered in this section. Using the numerical methods outlined in section 4, we hope to analyse the MRI stability condition for a disc with an equilibrium bending magnetic field ($B_{x0} \neq 0$). \\

For small magnetic field strengths, the question is significantly more complicated. The value of $B_{min}$ is related to the non-ideal behaviour of the disc, and therefore is beyond the scope of this analysis. It is unclear at this stage what effects MRI turbulence would have on disc structure or warp propagation.\\

\subsection{Comparison with hydrodynamic warped discs}

It is instructive to consider briefly how the magnetised warped disc considered here relates to the hydrodynamic warped disc discussed in \cite{2013MNRAS.433.2403O}.     \\

The distinctive characteristics of a hydrodynamic warped disc are largely due to the unbalanced pressure gradient in the radial equation of motion due to the warped geometry. Internal flows therefore must exist in a warped hydrodynamic disc. More specifically, linear flows were shown to be valid solutions. These flows will be resonantly driven for a Keplerian disc owing to the coincidence of the epicyclic and orbital frequencies. This leads to the rapid diffusion of the warp at small viscosities.\\

In contrast to a hydrodynamic disc, such simple linear flows are no longer solutions for the internal disc structure of a magnetised warped disc. The presence of the magnetic field adds some stiffness to the fluid, detuning the epicyclic frequency from the orbital frequency and thereby lifting the resonance. In more detail, it was found that there are a number of fast and slow Alfv\'{e}nic-epicyclic modes present with no analogue in the hydrodynamic disc. These modes may resonate if their natural frequency coincides with the orbital frequency, effectively driving a resonant response.\\

In an ideal inviscid disc the only non-zero internal torque coefficient is $Q_3$. The properties of $Q_3$ may be dramatically different from those of the corresponding hydrodynamic problem, reflecting the far richer structure of the magnetised warped disc. It is also expected that due to the detuning of the epicyclic and orbital frequencies in a warped magnetised disc, the anomalously large values of $Q_2$ found in the hydrodynamic Keplerian case will be reduced. \\

At progressively weaker field strengths (see Figure 4) the difference between the MHD solution and the corresponding hydrodynamic solution becomes negligible except when the disc is very close to an Alfv\'{e}nic-epicyclic resonance. Near these resonances the internal structure is very sensitive to small changes in parameters and the magnetohydrodynamic perturbations may have a significant effect on disc structure. This is of some physical relevance as a hydrodynamic Keplerian disc corresponds to one such resonance, and consequently may be sensitive to the introduction of a relatively weak magnetic field. Further investigation is required to probe the weak field limit and the effects of the MRI or a finite viscosity.

\section{Numerical solutions}

\subsection{Solution via numerical ODE solver}

In the previous section we described in detail a semi-analytical solution to the ideal MHD equations of a warped disc without mass outflow in the specific case where, at large distances, the external magnetic field is perpendicular to the disc surface. In this section we outline a numerical method for solving for the internal structure of an accretion disc. There are two motivations for this. Firstly, this provides a check on the semi-analytical results outlined in the previous section. Secondly, it would allow us to solve for the structure of the disc for a more general equilibrium magnetic field configuration and with more general boundary conditions. The method developed is outlined below.\\

	In section 3.2.2, we derived a set of six first-order ordinary differential equations for the six first-order variables; these are given by equations \eqref{eqlinstart}--\eqref{eqlinend}. The six necessary boundary conditions were also given in equations \eqref{bcstart}--\eqref{bcend}. In section 3.3.1 we derived three first-order equations for structure of the unwarped equilibrium disc, along with the three associated boundary conditions.\\
	
	Owing to symmetry considerations our range of integration may be restricted to $0<z'<z_{h}$, where $z_h$ is  the effective upper surface of the disc and is some number of scale heights above the midplane. $z_h$ was chosen to be small enough such that the numerical methods described remain feasible, yet large enough to be considered exterior to the disc for all practical purposes. All results were tested at various values of $z_h$ to ensure that its value did not have a significant impact.\\
	
	The boundary conditions were satisfied via the shooting method. The nine variables were integrated from the midplane to the upper surface of the disc. Four of the nine boundary conditions fix the value of a variable at the midplane, requiring the value of the five remaining variables at the midplane to be guessed. The Newton-Raphson method was employed to refine these five initial value guesses and satisfy the five boundary conditions at the upper surface. All results were found to be insensitive to subsequent iterations of the Newton-Raphson method and all boundary conditions were satisfied to an acceptable degree of error, leading to the conclusion that the Newton-Raphson method had converged on a valid solution.\\
	
	There are four free parameters in this numerical system. One is $B_{z'}$, a constant which determines the relative strength of the magnetic field in the system. The remaining three are the boundary conditions on the magnetic field which specify the connection between the magnetic field within the disc region and the magnetic field of the disc exterior. The first is $B_{x0}(z_h)$, determining the angle of the equilibrium poloidal field to the disc surface. The remaining two are $B_{x1}(z_h)$ and $B_{y1}(z_h)$, the first-order radial and toroidal magnetic fields external to the disc. These four free parameters must be specified to solve for the internal structure of the disc.\\
	
	As discussed in section 3.3.2, one of the boundary conditions on this system of equations was the imposition of a rigid lid corresponding to equation \eqref{bcend}. Based on a far-field analysis of the linear MHD equations, it was found that the results in the high-density region of the disc should be insensitive to the exact value of the outflow velocity. This was tested by varying the outflow velocity at the upper surface. We confirmed that the rigid lid had little effect on our solutions.\\
	
	All numerical results found via this method were consistent with the results of the semi-analytical calculation described in section 4 to a reasonable degree of accuracy.\\

 \subsection{The corrugated sheet model}
 
	As an alternative to the approach outlined here, one may also consider a standard shearing box model within which all perturbed variables vary sinusoidally with radius at a specified wavenumber $k$ (giving a thin disc the appearance of a corrugated sheet). For very small $k$ this corresponds exactly to the model presented above, where the warping is assumed to be independent of the radius. A numerical setup similar to the Runge-Kutta ODE solver outlined above was used for this corrugated sheet model. This model is informative because it allows us to investigate the dispersion relation $\omega(k)$ of radial bending waves. While the details of this model will be discussed in an upcoming paper, there are two results highly relevant to the work presented here.\\
	
	Firstly, the torque coefficient $Q_3$ can be found from the quadratic dispersion relation of the normal modes of the unwarped magnetised disc. In an unwarped disk there exists a normal mode with frequency $\omega(0) =\Omega$ (where $\Omega$ is the orbital frequency) corresponding to vertical oscillations of the disk. For small but non-zero radial wavenumber $k$, the frequency of this mode is related to $Q_3$ via \\
	
	\begin{equation}
	\omega(k) = \Omega + Q_3 \frac{c_s ^2}{\Omega} k^2.
	\end{equation}
It was found that the torque coefficients $Q_3$ obtained by this numerical method were consistent with the semi-analytical calculations presented in section 4, providing a valuable check on our results.\\
	
	Secondly, $\omega(k)$ at the critical points in the $(B_{z'},q)$ plane corresponding to Alfv\'{e}nic-epicyclic resonances were investigated. We found that the warp dispersion relation transitions from quadratic to linear near these critical points. This transition implies a qualitative change in the disc behaviour associated with enhanced warp propagation at these points. 
This result, described in more detail in the relevant upcoming paper, perhaps most clearly illustrates the physical interpretation of a divergent torque coefficient $Q_3$.

\section{Summary and discussion}

Using the warped shearing box framework set out by \cite{2013MNRAS.433.2403O}, the local ideal MHD equations were found for warped magnetised thin discs. These equations were expanded to terms linear in the warping parameter $\psi$. While a general analysis of the solutions to these equations is forthcoming, the case of an external magnetic field oriented normal to the disc surface is relatively simple and can be solved semi-analytically via a spectral method. The results of this analysis were verified by the use of a numerical ODE solver. Many of the distinct characteristics of warped magnetised discs may be illuminated by this rather simple model.\\

In a hydrodynamic disc it was found that the coincidence of the epicyclic and orbital frequencies causes a resonance in thin Keplerian accretion disc models. This resonance results in fast internal flows and torques acting within the disc, resulting in the rapid propagation of the warp.\\
 
	 In the presence of a magnetic field this resonance is removed. The magnetic tension adds a stiffness to the epicyclic oscillations, detuning the epicyclic frequency from the orbital frequency. Therefore at low viscosities the presence of even a relatively weak magnetic field may dramatically alter the internal structure of a Keplerian or very nearly Keplerian warped disc and evolution of the warp. \\
	 
	 In magnetised warped discs there exists a series of normal modes which we have called Alfv\'{e}nic-epicyclic modes. These are the normal modes of the magnetised accretion disc, each with a fixed vertical structure and phase relationship between the density, velocity and magnetic field perturbations. For a vanishing magnetic field, the fast Alfv\'{e}nic-epicyclic modes all oscillate at the orbital frequency while the slow modes are static ($\omega=0$), thereby recovering the results found by \cite{2013MNRAS.433.2403O}. The frequencies of the Alfv\'{e}nic-epicyclic modes depend on the shear rate $q$, the magnetic field strength and the inclination of the equilibrium magnetic field. \\
	 
	The warped geometry of the accretion disc creates a pressure gradient in the radial direction that acts as an inhomogeneous forcing term. This system is consequently analogous to a forced oscillator. When the frequency of an Alfv\'{e}nic-epicyclic mode coincides with the orbital frequency, that particular mode is resonantly forced resulting in large internal torques and rapid warp propagation. Warp propagation in magnetised discs consequently has a surprisingly subtle and rich dependence on the magnetic field strength and shear rate. This mechanism may be critical to understanding and modelling warp propagation in magnetised discs.\\
	
	In this paper we have assumed that there is no jet outflow from the disc. The effect of disc warping on jet outflows is not well understood and could be studied using the framework set out in this paper. We also hope to use the methods developed in this paper to investigate accretion discs with a wider variety of magnetic boundary conditions corresponding to more general magnetic field orientations.\\
	
	The solutions investigated in the latter half of this paper have neglected the MRI. Consequently, the results of those sections are expected to be of greatest validity in highly magnetised discs($\beta \approx 1$). The MRI, and especially the interplay between the MRI and the warped magnetised accretion disc dynamics outlined in this paper, could also be investigated using the formalism developed here. \\

This work was funded by the Science and Technology Facilities Council (STFC) through grant ST/L000636/1.

 {}

\begin{thebibliography}{1}
 
\bibitem[\protect\citeauthoryear{Agapitou, Papaloizou, \& Terquem}{1997}]{1997MNRAS.292..631A} Agapitou V., Papaloizou J.~C.~B., Terquem C., 1997, MNRAS, 292, 631  

\bibitem[\protect\citeauthoryear{Balbus \& Hawley}{1991}]{1991ApJ...376..214B} Balbus S.~A., Hawley J.~F., 1991, ApJ, 376, 214 

\bibitem[\protect\citeauthoryear{Blandford \& Payne}{1982}]{1982MNRAS.199..883B} Blandford R.~D., Payne D.~G., 1982, MNRAS, 199, 883 

\bibitem[\protect\citeauthoryear{Bardeen \& Petterson}{1975}]{1975ApJ...195L..65B} Bardeen J.~M., Petterson J.~A., 1975, ApJ, 195, L65 


\bibitem[\protect\citeauthoryear{Campbell}{2010}]{2010MNRAS.401..177C} Campbell C.~G., 2010, MNRAS, 401, 177 

\bibitem[\protect\citeauthoryear{Cao \& Spruit}{2013}]{2013ApJ...765..149C} Cao X., Spruit H.~C., 2013, ApJ, 765, 149 



\bibitem[\protect\citeauthoryear{Fragile et al.}{2007}]{2007ApJ...668..417F} Fragile P.~C., Blaes O.~M., Anninos P., Salmonson J.~D., 2007, ApJ, 668, 417 

\bibitem[\protect\citeauthoryear{Gammie \& Balbus}{1994}]{1994MNRAS.270..138G} Gammie C.~F., Balbus S.~A., 1994, MNRAS, 270, 138 


\bibitem[\protect\citeauthoryear{Guilet \& Ogilvie}{2012}]{2012MNRAS.424.2097G} Guilet J., Ogilvie G.~I., 2012, MNRAS, 424, 2097 


\bibitem[\protect\citeauthoryear{Hatchett, Begelman, \& Sarazin}{1981}]{1981ApJ...247..677H} Hatchett S.~P., Begelman M.~C., Sarazin C.~L., 1981, ApJ, 247, 677 

\bibitem[\protect\citeauthoryear{Krolik \& Hawley}{2015}]{2015ApJ...806..141K} Krolik J.~H., Hawley J.~F., 2015, ApJ, 806, 141


\bibitem[\protect\citeauthoryear{Lai}{1999}]{1999ApJ...524.1030L} Lai D., 1999, ApJ, 524, 1030 

\bibitem[\protect\citeauthoryear{Latter, Fromang, \& Gressel}{2010}]{2010MNRAS.406..848L} Latter H.~N., Fromang S., Gressel O., 2010, MNRAS, 406, 848 


\bibitem[\protect\citeauthoryear{Lewis, Bate, \& Price}{2015}]{2015MNRAS.451..288L} Lewis B.~T., Bate M.~R., Price D.~J., 2015, MNRAS, 451, 288 


\bibitem[\protect\citeauthoryear{Lewis \& Bate}{2017}]{2017MNRAS.467.3324L} Lewis B.~T., Bate M.~R., 2017, MNRAS, 467, 3324 


\bibitem[\protect\citeauthoryear{Lovelace \& Romanova}{2014}]{2014EPJWC..6405003L} Lovelace R.~V.~E., Romanova M.~M., 2014, EPJWC, 64, 05003 

\bibitem[\protect\citeauthoryear{Miyoshi et al.}{1995}]{1995Natur.373..127M} Miyoshi M., Moran J., Herrnstein J., Greenhill L., Nakai N., Diamond P., Inoue M., 1995, Natur, 373, 127 

\bibitem[\protect\citeauthoryear{Moll}{2012}]{2012A&A...548A..76M} Moll R., 2012, A\&A, 548, A76 


\bibitem[\protect\citeauthoryear{Narayan, Igumenshchev, \& Abramowicz}{2003}]{2003PASJ...55L..69N} Narayan R., Igumenshchev I.~V., Abramowicz M.~A., 2003, PASJ, 55, L69 

\bibitem[\protect\citeauthoryear{Nealon et al.}{2016}]{2016MNRAS.455L..62N} Nealon R., Nixon C., Price D.~J., King A., 2016, MNRAS, 455, L62 

\bibitem[\protect\citeauthoryear{Ogilvie}{1997}]{1997MNRAS.288...63O} Ogilvie G.~I., 1997, MNRAS, 288, 63 

\bibitem[\protect\citeauthoryear{Ogilvie}{1999}]{1999MNRAS.304..557O} Ogilvie G.~I., 1999, MNRAS, 304, 557 


\bibitem[\protect\citeauthoryear{Ogilvie}{2012}]{2012MNRAS.423.1318O} Ogilvie G.~I., 2012, MNRAS, 423, 1318 


\bibitem[\protect\citeauthoryear{Ogilvie \& Latter}{2013a}]{2013MNRAS.433.2403O} Ogilvie G.~I., Latter H.~N., 2013, MNRAS, 433, 2403 

\bibitem[\protect\citeauthoryear{Ogilvie \& Latter}{2013b}]{2013MNRAS.433.2420O} Ogilvie G.~I., Latter H.~N., 2013, MNRAS, 433, 2420 

\bibitem[\protect\citeauthoryear{Ogilvie \& Livio}{2001}]{2001ApJ...553..158O} Ogilvie G.~I., Livio M., 2001, ApJ, 553, 158 

\bibitem[\protect\citeauthoryear{Papaloizou \& Pringle}{1983}]{1983MNRAS.202.1181P} Papaloizou J.~C.~B., Pringle J.~E., 1983, MNRAS, 202, 1181 

\bibitem[\protect\citeauthoryear{Papaloizou \& Lin}{1995}]{1995ApJ...438..841P} Papaloizou J.~C.~B., Lin D.~N.~C., 1995, ApJ, 438, 841 



\bibitem[\protect\citeauthoryear{Sheikhnezami \& Fendt}{2015}]{2015ApJ...814..113S} Sheikhnezami S., Fendt C., 2015, ApJ, 814, 113 


\bibitem[\protect\citeauthoryear{Sorathia, Krolik, \& Hawley}{2013}]{2013ApJ...777...21S} Sorathia K.~A., Krolik J.~H., Hawley J.~F., 2013, ApJ, 777, 21 

\bibitem[\protect\citeauthoryear{Sorathia, Krolik, \& Hawley}{2013}]{2013ApJ...768..133S} Sorathia K.~A., Krolik J.~H., Hawley J.~F., 2013, ApJ, 768, 133 

\bibitem[\protect\citeauthoryear{Terquem \& Papaloizou}{2000}]{2000A&A...360.1031T} Terquem C., Papaloizou J.~C.~B., 2000, A\&A, 360, 1031 


\bibitem[\protect\citeauthoryear{Zhuravlev et al.}{2014}]{2014ApJ...796..104Z} Zhuravlev V.~V., Ivanov P.~B., Fragile P.~C., Morales Teixeira D., 2014, ApJ, 796, 104 



\end{thebibliography}
\end{document}